\documentclass[11pt]{article}

\newcommand{\slpart}{\mbox{$\partial \hspace{-0.50em}/$}}

\newcommand{\sln}{\mbox{$n \hspace{-0.50em}/$}}
\newcommand{\slP}{\mbox{$P \hspace{-0.65em}/$}}
\newcommand{\slQ}{\mbox{$Q \hspace{-0.65em}/$}}

\usepackage{amssymb}
\usepackage{amsmath}
\usepackage{amsfonts}
\usepackage{ulem}
\usepackage{axodraw}
\usepackage{simplewick}
\usepackage[toc,page]{appendix}
\usepackage{hyperref}

\title{Pion-Nucleon Scattering in Kadyshevsky Formalism: I Meson Exchange Sector}
\author{J.W.Wagenaar and T.A.Rijken}

\begin{document}
\allowdisplaybreaks \maketitle

\abstract{In a series of two papers we present the theoretical
results of $\pi N$/meson-baryon scattering in the Kadyshevsky
formalism. In this paper the results are given for meson exchange
diagrams. On the formal side we show, by means of an example, how
general couplings, i.e. couplings containing multiple derivatives
and/or higher spin fields, should be treated. We do this by
introducing and applying the Takahashi-Umezawa and the
Gross-Jackiw method. For practical purposes we introduce the
$\bar{P}$ method. We also show how the Takashashi-Umezawa method
can be derived using the theory of Bogoliubov and collaborators
and the Gross-Jackiw method is also used to study the
$n$-dependence of the Kadyshevsky integral equation. Last but not
least we present the second quantization procedure of the quasi
particle in Kadyshevsky formalism.}

\section{Introduction}

Over the years the Nijmegen group has constructed very successful
baryon-baryon models (NN and YN). As for instance in
\cite{nagels78} and \cite{maessen89} soft-core One-Boson-Exchange
NN and YN models are constructed based on Regge-pole theory. The
models are linked via $SU_f(3)$ symmetry in order to have more
control on the parameters.

Based on the same ideas, the Nijmegen group recently broadened its
horizon by also including meson-baryon models \cite{henk1}. Here,
a simultaneous $\pi N$ and $K^+ N$ model is constructed using
one-meson and one-baryon exchange potentials.

This work is presented in two articles, referred to as paper I
(this paper) and paper II \cite{JWT2}, and can be regarded as an
extension of \cite{henk1}, since we also consider meson-baryon
scattering or pion-nucleon, more specifically. The reason for
considering pion-nucleon scattering is, besides the interest in
its own, that there is a large amount of experimental data. Using
the aforementioned $SU_f(3)$ symmetry the extension to other
meson-baryon systems is easily made. Last but not least we would
like to mention the connection to photo/electro-production models.

Compared to \cite{henk1} our focus is more on the theoretical
background. For instance we formally include what is called "pair
suppression", whereas this was assumed in \cite{henk1}. Pair
suppression comes down to the suppression of negative energy
contributions. For the first time, at least to our knowledge, we
incorporate pair suppression in a covariant and frame independent
way. This may also be interesting for relativistic many body
theories. The details of the formal incorporation of pair
suppression are discussed in paper II.

In order to have this covariant and frame independent pair
suppression, we use the Kadyshevsky formalism
\cite{Kad64,Kad67,Kad68,Kad70}. This formalism is equivalent to
Feynman formalism, since it can be derived from the same S-matrix
formula. It covariantly, though frame dependently \footnote{By
frame dependent we mean: dependent on a vector $n^\mu$.},
separates positive and negative energy contributions. Generally,
the number of diagrams increases: $1\rightarrow n!$ at order $n$
as in old-fashioned perturbation theory. Contrary to the Feynman
formalism all particles in the Kadyshevsky formalism remain on
their mass shell, at the cost of the introduction of an extra
quasi particle, which carries four momentum only. A second
quantization formalism of this quasi field is presented in
appendix \ref{secquant}. An other advantage of the Kadyshevsky
formalism is that it brings about a three dimensional
Lippmann-Schwinger type of integral equation \cite{Kad70}, whereas
a three dimensional integral equation was achieved in \cite{henk1}
only after approximations of the Bethe-Salpeter equation
\cite{betsalp}. We study the $n$-dependence of the Kadyshevsky
integral equation with tree level amplitudes as input in section
\ref{ndepkad}. As compared to the original Kadyshevsky rules we
use a slightly different version, introduced and discussed in
appendix \ref{kadrules}.

Couplings containing derivatives and higher spin fields may cause
differences and problems as far as the results in the Kadyshevsky
formalism and the Feynman formalism are concerned. This is
discussed in section \ref{ex1} by means of an example of
simplified vector meson exchange. After a second glance the
results in both formalisms are the same, however, they contain
extra frame dependent contact terms. Two methods are introduced
and applied, which discuss a second source of extra terms: the
Takahashi-Umezawa (TU) \cite{Tak53a,Tak53b,Ume56} and the
Gross-Jackiw (GJ) \cite{Gross69} method. The extra terms coming
from this second source cancel the former ones exactly. Both
formalisms, however, yield the same results. With the use of (one
of) these methods the final results for the S-matrix or amplitude
are covariant and frame independent ($n$-independent). In section
\ref{Pbar} we introduce and discusse the $\bar{P}$-method, which
is quite useful for practical purposes. We derive the TU method
from the BMP \cite{BMP58,BS59,BLT75} theory in appendix \ref{BMP}
and in light of this TU method we make some remarks about the Haag
theorem \cite{haag55} in appendix \ref{haag}.

Although we already discussed some content, this paper is
organized as follows: we start in section \ref{MBkin} with some
meson-baryon scattering kinematics in Kadyshevsky formalism
together with the discussion of the $n$-dependence of the integral
equation. We start the application of the Kadyshevsky formalism to
the $\pi N$ system by first discussing the ingredients of the
model in section \ref{OBE}. The meson exchange amplitudes are
calculated in section \ref{mesonex}, which contains the results
for equal initial and final states. For the results for general
meson-baryon initial and final states we refer to appendix A of
paper II. For the results for baryon exchange we refer to paper II
as well. As mentioned before section \ref{mesonex} also contains
the discussion of how general couplings, i.e. couplings containing
multiple derivatives and/or higher spin fields, should be treated
in the Kadyshevsky formalism.

\section{Meson-Baryon Scattering Kinematics}\label{MBkin}

We consider the pion-nucleon or more general the meson-baryon
reactions
\begin{equation}
  M_i(q)+B_i(p,s) \rightarrow M_f(q')+B_f(p',s')\ .\label{mbkin1}
\end{equation}
where $M$ stands for a meson and $B$ is a baryon. For the four
momentum of the baryons and mesons we take, respectively
\begin{eqnarray}
 p^\mu_{c}=\left(E_{c},{\bf p}_{c}\right)\quad&,&where\quad E_{c}=\sqrt{{\bf
 p}_{c}^{2}+M_{c}^{2}}\ ,\nonumber\\*
 q^\mu_{c}=\left(\mathcal{E}_{c},{\bf q}_{c}\right)\quad&,&where\quad \mathcal{E}_{c}=\sqrt{{\bf
 q}_{c}^{2}+m_{c}^{2}}\ .\label{mbkin2}
\end{eqnarray}
Here, $c$ stands for either the initial state $i$ or the final
state $f$. In some cases we find it useful to use the definitions
\eqref{mbkin2} for the intermediate meson-baryon states $n$.

Using the Kadyshevsky formalism (appendix \ref{kadrules}) and
especially the second quantization procedure (appendix
\ref{secquant}) external quasi particles may occur with initial
and final state momenta $n\kappa$ and $n\kappa'$, respectively.
Therefore, the usual overall four-momentum conservation is
generally replaced by
\begin{equation}
 p+q+\kappa\ n = p'+q'+ \kappa'\ n\ .\label{mbkin3}
\end{equation}
As \eqref{mbkin3} and \eqref{mbkin1} make clear a "prime" notation
is used to indicate final state momenta; no prime means initial
state momenta. We will maintain this notation (also for the
energies) throughout these articles, unless indicated otherwise.

Furthermore we find it useful to introduce the Mandelstam
variables in the Kadyshevsky formalism
\begin{eqnarray}
 s_{pq} &=& (p+q)^2\ \ , \ \ s_{p'q'}=(p'+q')^2\ , \nonumber\\
 t_{p'p}&=& (p'-p)^2\ \ , \ \ t_{q'q}=(q'-q)^2\ , \nonumber\\
 u_{p'q}&=& (p'-q)^2\ \ , \ \ u_{pq'}=(p-q')^2\ ,\label{mbkin4}
\end{eqnarray}
where $s_{pq}$ and $s_{p'q'}$ etc., are only identical for
$\kappa'=\kappa=0$. These Mandelstam variables satisfy the
relation
\begin{equation}
 2 \sqrt{s_{p'q'}s_{pq}} + t_{p'p} + t_{q'q} + u_{pq'} + u_{p'q} =
 2 \left(M_f^2 + M_i^2 + m_f^2 + m_i^2 \right)\ .\label{mbkin5}
\end{equation}
\\
The total and relative four-momenta of the initial, final, and
intermediate channel $(c=i,f,n)$ are defined by
\begin{eqnarray}
 P_c &=& p_c + q_c\ ,\ k_c = \mu_{c,2}\ p_c - \mu_{c,1}\ q_c\ ,
 \label{mbkin6}
\end{eqnarray}
where the weights satisfy $\mu_{c,1}+\mu_{c,2}=1$. We choose the
weights to be
\begin{eqnarray}
 \mu_{c,1}&=&\frac{E_c}{E_c+\mathcal{E}_c}\ ,\nonumber\\*
 \mu_{c,2}&=&\frac{\mathcal{E}_c}{E_c+\mathcal{E}_c}\ .\label{mbkin7}
\end{eqnarray}
Since in the Kadyshevsky formalism all particles are on their mass
shell, the choice \eqref{mbkin7} means that always $k_c^0=0$.

In the center-of-mass (CM) system ${\bf p}=-{\bf q}$ and ${\bf
p'}=-{\bf q'}$, therefore
\begin{eqnarray}
 P_i = (W,{\bf 0})\ , && P_f =(W',{\bf 0})\ ,\nonumber\\
 k_i = (0,{\bf p})\ , && k_f = (0,{\bf p'})\ .\label{mbkin8}
\end{eqnarray}
where $W= E+{\cal E}$ and $W'= E'+{\cal E'}$. Furthermore we take
$n^\mu=(1,{\bf 0})$.

Also we take as the scattering plane the xz-plane, where the
3-momentum of the initial baryon is oriented in the positive
z-direction.

In the CM system the unpolarized differential cross section is
defined to be
\begin{equation}
 \left(\frac{d\sigma}{d\Omega}\right)_{CM}
 =\frac{|{\bf p}'|}{2|{\bf p}|}\sum\left|\frac{M_{fi}}{8\pi\sqrt{s}}\right|^2\
 ,\label{mbkin9}
\end{equation}
where the amplitude $M_{fi}$ is defined in appendix \ref{kadrules}
and the sum is over the spin components of the final baryon.

To generate amplitudes at all orders we use the Kadyshevsky
integral equation in the CM system
\begin{eqnarray}
 M(W'\,{\bf p'};W\,{\bf p})
&=&
 M_{00}^{irr}(W'\,{\bf p'};W\,{\bf p})
 +\int d^3k_n\,M_{0\kappa}^{irr}(W'\,{\bf p'};W_n\,{\bf k}_n)\nonumber\\
&&
 \times\frac{1}{(2\pi)^3}\ \frac{1}{4\mathcal{E}_{n}E_{n}}\
 \frac{1}{\sqrt{s}-\sqrt{s_n}+i\varepsilon}\
 M_{\kappa0}(W_n\,{\bf k}_n;W\,{\bf p})\ .\label{KIE7}
\end{eqnarray}
Although there are still $\kappa$-labels in \eqref{KIE7}, they're
fixed at $\kappa=P_i^0-P_n^0$. Also we have included the spinors
of the projection operator of the fermion propagator
\begin{eqnarray}
 S^{(+)}(p_n)
&=&
 \Lambda^{(1/2)}(p_n)\ \theta(p_n^0)\delta(p_n^2-M^2)\ ,\nonumber\\
&=&
 \sum_{s_n}
 u(p_ns_n)\bar{u}(p_ns_n)\ \theta(p_n^0)\delta(p_n^2-M^2)\
 ,\label{KIE2a}
\end{eqnarray}
in the amplitudes $M_{0\kappa}(p'q';p_nq_n)$ and
$M_{\kappa0}(p_nq_n;pq)$.

We have put the intermediate negative energy states
($\Delta^{(-)}(x-y;m_\pi^2)$ and $S^{(-)}(x-y;M_N^2)$) in
$M^{irr}_{\kappa\kappa'}$, but in principle they could also
participate in the integral equation. However, using pair
suppression in the way we do in paper II, these terms vanish.

\subsection{$n$-independence of Kadyshevsky Integral Equation}
\label{ndepkad}

When generating Kadyshevsky diagrams to random order using the
Kadyshevsky integral equation, the (full) amplitude is identical
to the one obtained in Feynman formalism when the external quasi
particle momenta are put to zero. It is therefore $n$-independent,
i.e. frame independent.

Since an approximation is used to solve the Kadyshevsky integral
equation, namely tree level diagrams as driving terms, it is not
clear whether the full amplitude remains to be $n$-independent
when the external quasi particle momenta are put to zero.

In examining the $n$-dependence of the amplitude we write the
Kadyshevsky integral equation schematically as
\begin{equation}
 M_{00} = M_{00}^{irr} + \int d\kappa\ M_{0\kappa}^{irr}\ G'_\kappa\ M_{\kappa0}\
 ,\label{KIE10}
\end{equation}
Since $n^2=1$, only variations in a space-like direction are
unrestricted, i.e. $n\cdot\delta n=0$ \cite{Gross69}. We therefore
introduce the projection operator
\begin{equation}
 P^{\alpha\beta} = g^{\alpha\beta} - n^\alpha n^\beta\ ,\label{KIE11}
\end{equation}
from which it follows that $n_\alpha P^{\alpha\beta}=0$. The
$n$-dependence of the amplitude can now be studied
\begin{eqnarray}
 P^{\alpha\beta}\frac{\partial}{\partial n^\beta} M_{00}
&=&
 P^{\alpha\beta}\frac{\partial M_{00}^{irr}}{\partial n^\beta}
 \nonumber\\
&&
 +P^{\alpha\beta}\int d\kappa\left[\frac{\partial M_{0\kappa}^{irr}}{\partial n^\beta}\
 G'_\kappa\ M_{\kappa0} + M_{0\kappa}^{irr}\ G'_\kappa\
 \frac{\partial M_{\kappa0}}{\partial n^\beta}\right]\
 .\qquad\quad\label{KIE12}
\end{eqnarray}
If both Kadyshevsky contributions are considered at second order
in $M_{00}$, then it is $n$-independent, since it yields the
Feynman expression. As far as the second term in \eqref{KIE12} is
concerned we observe the following
\begin{equation}
 \frac{\partial M_{0\kappa}^{irr}}{\partial n^\beta} \propto \kappa f(\kappa)\ \ ,\ \
 \frac{\partial M_{\kappa0}}{\partial n^\beta} \propto \kappa g(\kappa)\ , \label{KIE13}
\end{equation}
where $f(\kappa)$ and $g(\kappa)$ are functions that do not
contain poles or zero's at $\kappa=0$. Therefore, the integral in
\eqref{KIE12} is of the form
\begin{equation}
 \int d\kappa\ \kappa\ h(\kappa) G'_\kappa\ .\label{KIE13a}
\end{equation}

\begin{figure}[Ht]
 \begin{center} \begin{picture}(200,120)(0,0)
 \SetPFont{Helvetica}{9}
 \SetScale{1.0} \SetWidth{0.2}

 \Line(0,20)(200,20)
 \Line(100,0)(100,120)
 \Text(145,10)[]{$\Re(\kappa)$}
 \Text(85,85)[]{$\Im(\kappa)$}

 \SetWidth{1.5}
 \ArrowLine(0,20)(80,20)
 \ArrowLine(120,20)(200,20)
 \Vertex(100,20){3}
 \Text(40,30)[]{\bf I}
 \Text(160,30)[]{\bf I}

 \ArrowArc(100,20)(100,0,180)
 \ArrowArcn(100,20)(20,-180,0)
 \DashArrowArcn(100,20)(20,0,180){4}
 \Text(110,110)[]{\bf II}
 \Text(110,50)[]{\bf III}
 \Vertex(15,12){3}
 \Vertex(185,12){3}

 \end{picture}
 \end{center}
 \caption{\sl Principle value integral}
\label{fig:Pintegral}
\end{figure}
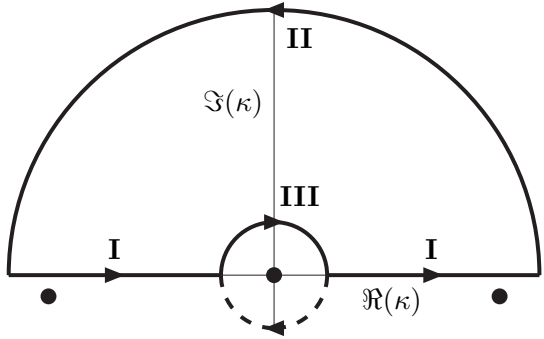

When performing the integral we decompose the $G'_\kappa$ as
follows
\begin{eqnarray}
 G'_\kappa\propto
 \frac{1}{\kappa+i\varepsilon}=P\frac{1}{\kappa}-i\pi\delta(\kappa)\
 .\label{KIE14}
\end{eqnarray}
As far as the $\delta(\kappa)$-part of \eqref{KIE14} is concerned
we immediately see that it gives zero when used in the integral
\eqref{KIE13a}. For the Principle valued integral, indicated in
figure \ref{fig:Pintegral} by {\bf I}, we close the integral by
connecting the end point ($\kappa=\pm\infty$) via a (huge)
semi-circle in the upper half, complex $\kappa$-plane (line {\bf
II} in figure \ref{fig:Pintegral}) and by connecting the points
around zero via a small semi circle also in the upper half plane
(line {\bf III} in figure \ref{fig:Pintegral}). Since every single
(tree level) amplitude is proportional to
$1/(\kappa+A+i\varepsilon)$, where $\kappa$ is related to the
momentum of the incoming or outgoing quasi particle and $A$ some
positive or negative number, the poles will always be in the lower
half plane and not within the contour. Therefore, the contour
integral is zero.

Since we have added integrals ({\bf II} and {\bf III} in figure
\ref{fig:Pintegral}) we need to know what their contributions are.
The easiest part is integral {\bf III}. Its contribution is half
the residue at $\kappa=0$ and since the only remaining integrand
part $h(\kappa)$ in \eqref{KIE13a} doesn't contain a pole at zero
it is zero.

If we want the contribution of integral {\bf II} to be zero, than
the integrand should at least be of order $O(\frac{1}{\kappa^2})$.
Unfortunately, this is not (always) the case as we will see in
sections \ref{mesonex} and paper II. To this end we introduce a
phenomenological "form factor"
\begin{eqnarray}
 F(\kappa)=\left(\frac{\Lambda^2_\kappa}
 {\Lambda^2_\kappa-\kappa^2-i\epsilon(\kappa)\varepsilon}\right)^{N_\kappa}\
 ,\label{KIE15}
\end{eqnarray}
where $\Lambda_\kappa$ is large and $N_\kappa$ is some positive
integer. In \eqref{KIE15} $\varepsilon$ is real, positive, though
small and $\epsilon(\kappa)=\theta(\kappa)-\theta(-\kappa)$.

The effect of the function $F(\kappa)$ \eqref{KIE15} on the
original integrand in \eqref{KIE13a} is little, since for large
$\Lambda_\kappa$ it is close to unity. However, including this
function in the integrand makes sure that it is at least of order
$O(\frac{1}{\kappa^2})$ so that integral {\bf II} gives zero
contribution. The $-i\epsilon(\kappa)\varepsilon$ part ensures
that there are now poles on or within the closed contour, since
they are always in the lower half plane (indicated by the dots in
figure \ref{fig:Pintegral}).

\section{Application: Pion-Nucleon Scattering}\label{OBE}

In the following sections we're going to apply the Kadyshevsky
formalism to the pion-nucleon system, although we present it in
such a way that it can easily be extended to other meson-baryon
systems. The isospin factors are not included in our treatment; we
are only concerned about the Lorentz and Dirac structure. For the
details about the isospin factors we refer to \cite{henk1}.

\begin{figure}[hbt]
 \begin{center} \begin{picture}(410,110)(0,0)


 \SetScale{1.0} \SetWidth{1.5}
 \ArrowLine(0,20)(55,30)
 \ArrowLine(55,30)(110,20)
 \DashArrowLine(0,100)(55,90){4}
 \DashArrowLine(55,90)(110,100){4}
 \Photon(55,30)(55,90){4}{3}
 \Vertex(55,30){4}
 \Vertex(55,90){4}
 \Vertex(55,60){2}

 \Text(55,0)[]{t : $f_0,\sigma,P,\rho$}

 \SetWidth{0.2}
 \ArrowLine(0,60)(55,60)
 \ArrowLine(55,60)(110,60)


 \SetScale{1.0} \SetWidth{1.5}
 \ArrowLine(150,20)(205,20)
 \Line(205,20)(205,100)
 \ArrowLine(205,100)(260,100)
 \DashArrowLine(150,100)(205,100){4}
 \DashArrowLine(205,20)(260,20){4}
 \Vertex(205,20){4}
 \Vertex(205,100){4}
 \Vertex(205,60){2}

 \Text(205,0)[]{u : $N,N^*,\Delta$}

 \SetWidth{0.2}
 \ArrowLine(150,60)(205,60)
 \ArrowLine(205,60)(260,60)


 \SetPFont{Helvetica}{9}
 \SetScale{1.0} \SetWidth{1.5}

 \ArrowLine(300,20)(330,60)
 \Line(330,60)(380,60)
 \ArrowLine(380,60)(410,20)
 \DashArrowLine(300,100)(330,60){4}
 \DashArrowLine(380,60)(410,100){4}
 \Vertex(330,60){4}
 \Vertex(380,60){4}
 \Vertex(355,60){2}

 \Text(355,0)[]{s : $N,N^*,\Delta$}

 \SetWidth{0.2}
 \ArrowLine(300,110)(355,60)
 \ArrowLine(355,60)(410,110)

\end{picture}
\end{center}
\caption{\sl Tree level amplitudes as input for integral equation.
 The inclusion of the quasi particle lines is schematically.
 Therefore, the diagrams represent either the $(a)$ or the
 $(b)$ diagram.}\label{fig:ingred}
\end{figure}
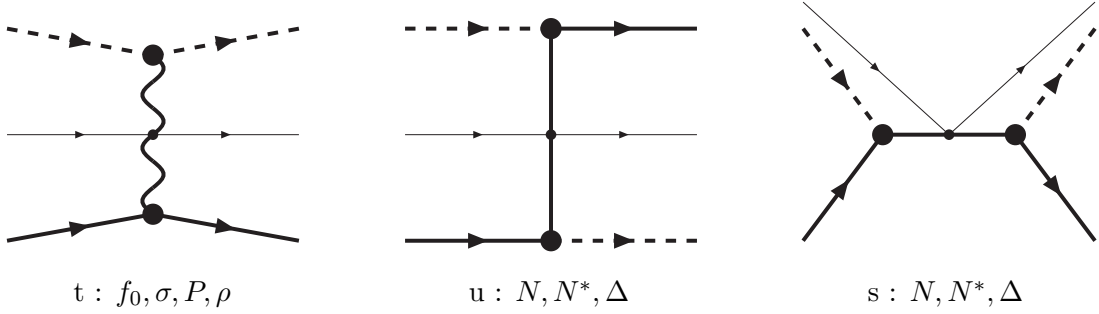

The ingredients of the model are tree level, exchange amplitudes
as mentioned before. These amplitudes serve as input for the
integral equation. Very similar to what is done in \cite{henk1} we
consider for the amplitudes the exchanged particles as in table
\ref{tab:ingred}.
\begin{table}
\begin{center}
\begin{tabular}{|c|c|}
  \hline
  Channel & Exchanged particle \\
  \hline
  t & $f_0,\sigma,P,\rho$  \\
  u & $N,N^*,S_{11},\Delta_{33}$ \\
  s & $N,N^*,S_{11},\Delta_{33}$ \\
  \hline
\end{tabular}
\end{center}
 \caption{\sl Exchanged particles in the various channels}\label{tab:ingred}
\end{table}
Graphically, this shown in figure \ref{fig:ingred}.
Contrary to \cite{henk1} we do not consider the exchange of the
tensor mesons, since their contribution is little. The inclusion
of the them can be regarded as an extension of this work.\\

\noindent For the description of the amplitudes we need the
interaction Lagrangians, which in our treatment always serve as
the starting points
\begin{itemize}
 \item \underline{Triple meson vertices}
 \begin{subequations}
 \begin{eqnarray}
  {\cal L}_{SPP}&=&g_{PPS}\,\phi_{P,a}\phi_{P,b}\cdot \phi_S\
  ,\label{ing1a}\\*
  {\cal L}_{VPP}&=&g_{VPP}\left(\phi_a i\overset{\leftrightarrow}{\partial}_{\mu}\phi_b
                   \right)\phi^{\mu}\ ,\label{ing1b}
 \end{eqnarray}
 \end{subequations}
 where $S,V,P$ stand for {\it scalar}, {\it vector} and {\it pseudo
 scalar} to indicate the various mesons. The indices $a,b$ are used
 to indicate the outgoing and incoming meson, respectively. For the
derivative $\overleftrightarrow{\partial_\mu} =
\overrightarrow{\partial_\mu} - \overleftarrow{\partial_\mu}$.
 \item \underline{Meson-baryon vertices}
 \begin{subequations}
 \begin{eqnarray}
  {\cal L}_{SNN}&=&g_{S}\,\bar{\psi}\psi\cdot \phi_S\ ,\label{ing2a}\\
  {\cal L}_{VNN}&=&g_{V}\,\bar{\psi}\gamma_{\mu}\psi\ \phi^{\mu}
                   -\frac{f_V}{2M_V}\ i\partial^{\mu}\left(\bar{\psi}\sigma_{\mu\nu}\psi\right)\cdot\phi^{\nu}
                   \ ,\label{ing2b}\\
  {\cal L}_{PV}&=&\frac{f_{PV}}{m_{\pi}}\,\bar{\psi}\gamma_5\gamma_\mu\psi\cdot\partial^\mu\phi_P\ ,\label{ing2c}\\
  {\cal L}_{V}&=&\frac{f_{V}}{m_{\pi}}\,\bar{\psi}\gamma_\mu\psi\cdot\partial^\mu\phi_P\ ,\label{ing2d}
 \end{eqnarray}
 \end{subequations}
 where $\sigma_{\mu\nu}=\frac{1}{2}\left[\gamma_\mu,\gamma_\nu\right]$. The
 coupling constants $f_V$ of \eqref{ing2b} and
 \eqref{ing2d} do not necessarily coincide.

 We have chosen \eqref{ing2b} in such a way that the vector meson
 couples to a current, which may contain a derivative. This is a
 bit different from \cite{henk1,Ver76}, where the derivative acts on the
 vector meson. In Feynman theory this does not make a difference,
 however it does in Kadyshevsky formalism, because of the presence
 of the quasi particles.

 Equation \eqref{ing2c} is used to describe the exchange
 ($u,s$-channel) of the nucleon and Roper ($N^*$) and \eqref{ing2d}
 is used for the $S_{11}$ exchange. This, because of their
 intrinsic parities. Note, that we could also have chosen the
 pseudo scalar and scalar couplings for these exchanges. However,
 since the interactions \eqref{ing2c} and \eqref{ing2d} are also
 used in \cite{henk1} and in chiral symmetry based models, we use
 these interactions.
 \item \underline{$\pi N\Delta_{33}$ vertex}
 \begin{eqnarray}
  {\cal L}_{\pi N\Delta}
 &=&
  g_{gi}\,\epsilon^{\mu\nu\alpha\beta}\left(\partial_{\mu}\bar{\Psi}_\nu\right)\gamma_5\gamma_{\alpha}
  \psi\left(\partial_{\beta}\phi\right)
  +g_{gi}\,\epsilon^{\mu\nu\alpha\beta}\bar{\psi}\gamma_5\gamma_{\alpha}
  \left(\partial_{\mu}\Psi_{\nu}\right)\left(\partial_{\beta}\phi\right)
  \ ,\label{ing3}
 \end{eqnarray}
 The use of this interaction Lagrangian differs from the
 one used in \cite{henk1}. We'll come back to this in paper II.
\end{itemize}

\noindent The meson exchange processes are discussed in section
\ref{mesonex}. As mentioned before the discussion of the baryon
exchange processes (including pair suppression) is postponed to
paper II. An other important ingredient of the model is the use of
form factors. We also postpone the discussion of them to paper II.

\section{Meson Exchange}\label{mesonex}

Here, we proceed with the discussion of the meson exchange
processes. We give the amplitudes for meson-baryon scattering or
pion-nucleon scattering, specifically, meaning that we take equal
initial and final states ($M_f=M_i=M$ and $m_f=m_i=m$, where $M$
and $m$ are the masses of the nucleon and pion, respectively). The
results for general meson-baryon initial and final states are
presented in appendix A of paper II.

\subsection{Scalar Meson Exchange}\label{sectionsme}

\begin{figure}[hbt]
 \begin{center} \begin{picture}(400,110)(0,0)
 \SetPFont{Helvetica}{9}
 \SetScale{1.0} \SetWidth{1.5}
 \DashArrowLine(50,90)(100,90){4}
 \DashArrowLine(100,90)(150,90){4}
 \ArrowLine(50,20)(100,20)
 \ArrowLine(100,20)(150,20)
 \Vertex(100,90){3}
 \Vertex(100,20){3}
 \DashArrowLine(100,90)(100,20){2}

 \Text(40,90)[]{$q$}
 \Text(40,20)[]{$p$}
 \Text(160,90)[]{$q'$}
 \Text(160,20)[]{$p'$}
 \Text( 90,55)[]{$P_a$}


 \SetWidth{0.2}
 \ArrowLine(50,105)(100,90)
 \ArrowArcn(65,55)(49.5,45,315)
 \ArrowLine(100,20)(150,05)
 \Text(40,105)[]{$\kappa$}
 \Text(160,05)[]{$\kappa'$}
 \Text(125,55)[]{$\kappa_1$}
 \PText(100,00)(0)[b]{(a)}

 \SetScale{1.0} \SetWidth{1.5}
 \DashArrowLine(250,90)(300,90){4}
 \DashArrowLine(300,90)(350,90){4}
 \ArrowLine(250,20)(300,20)
 \ArrowLine(300,20)(350,20)
 \Vertex(300,90){3}
 \Vertex(300,20){3}
 \DashArrowLine(300,20)(300,90){2}

 \Text(240,90)[]{$q$}
 \Text(240,20)[]{$p$}
 \Text(360,90)[]{$q'$}
 \Text(360,20)[]{$p'$}
 \Text(290,55)[]{$P_b$}


 \SetWidth{0.2}
 \ArrowLine(250,05)(300,20)
 \ArrowArc (265,55)(49.5,315,45)
 \ArrowLine(300,90)(350,105)
 \Text(240,05)[]{$\kappa$}
 \Text(360,105)[]{$\kappa'$}
 \Text(325,55)[]{$\kappa_1$}
 \PText(300,00)(0)[b]{(b)}

\end{picture}
\end{center}
\caption{\sl Scalar meson exchange}\label{fig:scalar}
\end{figure}
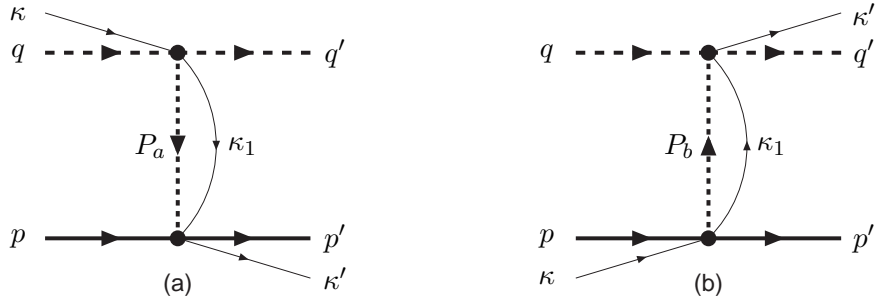

For the description of the scalar meson exchange processes at tree
level, graphically shown in figure \ref{fig:scalar}, we use the
interaction Lagrangians \eqref{ing1a} and \eqref{ing2a}, which
lead to the vertices
\begin{eqnarray}
 \Gamma_{PPS}&=&g_{PSS}\ ,\nonumber\\
 \Gamma_{S}&=&g_{S}\ ,\label{sc1}
\end{eqnarray}
using $\mathcal{L}_I=-\mathcal{H}_I\rightarrow-\Gamma$. For the
appropriate propagator we use the first line of \eqref{wick6}.

Applying the Kadyshevsky rules as discussed in appendix
\ref{kadrules}, the amplitudes read
\begin{eqnarray}
 M^{a,b}_{\kappa'\kappa}
&=&
 g_{PSS}g_{S}\,\int\frac{d\kappa_1}{\kappa_1+i\varepsilon}\left[\bar{u}(p's')u(ps)\right]
 \theta(P^0_{a,b})\delta(P^2_{a,b}-M_S^2)\ ,\label{sc2}
\end{eqnarray}
where $P_{a,b}=\pm\Delta_t+\frac{1}{2}(\kappa'+\kappa)-n\kappa_1$
(here $a$ corresponds to the $+$ sign and $b$ to the $-$ sign) and
$\Delta_t=\frac{1}{2}(p'-p-q'+q)$. For the $\kappa_1$ integration
we consider the $\delta$-function in \eqref{sc2}
\begin{eqnarray}
 (a):
&&
 \delta(P_a^2-M_S^2)=\frac{1}{|\kappa_1^{+}-\kappa_1^{-}|}
 \left(\delta(\kappa_1-\kappa_1^{+})+\delta(\kappa_1-\kappa_1^{-})\right)\ ,\nonumber\\
&&
 \phantom{\delta(P_a^2-M_S^2)=}\kappa_1^{\pm}
 =\Delta_t\cdot n+\frac{1}{2}\left(\kappa'+\kappa\right)\pm A_t\ ,\nonumber\\
 (b):
&&
 \delta(P_b^2-M_S^2)=\frac{1}{|\kappa_1^{+}-\kappa_1^{-}|}
 \left(\delta(\kappa_1-\kappa_1^{+})+\delta(\kappa_1-\kappa_1^{-})\right)\ ,\nonumber\\
&&
 \phantom{\delta(P_a^2-M_S^2)=}\kappa_1^{\pm}=-\Delta_t\cdot n+\frac{1}{2}\left(\kappa'+\kappa\right)\pm
 A_t\ ,\label{sc2a}
\end{eqnarray}
where $A_t=\sqrt{(n\cdot\Delta_t)^2-\Delta_t^2+M_S^2}$. In both
cases $\theta(P^0_{a,b})$ selects the $\kappa_1^{-}$ solution.
Therefore,
\begin{eqnarray}
 P_a&=&\Delta_t-\left(\Delta_t\cdot n\right)n+A_t n\ ,\nonumber\\
 P_b&=&-\Delta_t+\left(\Delta_t\cdot n\right)n+A_t n\ .\label{sc2b}
\end{eqnarray}
With these expression we find for the amplitudes
\begin{eqnarray}
  M^{(a)}_{\kappa'\kappa}
&=&
 g_{PSS}g_{S}\,\left[\bar{u}(p's')u(ps)\right]\frac{1}{2A_t}\cdot
 \frac{1}{\Delta_t\cdot n+\bar{\kappa}-A_t+i\varepsilon}\ ,\nonumber\\
 M^{(b)}_{\kappa'\kappa}
&=&
 g_{PSS}g_{S}\,\left[\bar{u}(p's')u(ps)\right] \frac{1}{2A_t}\cdot
 \frac{1}{-\Delta_t\cdot n+\bar{\kappa}-A_t+i\varepsilon}\ ,\qquad\label{sc3}
\end{eqnarray}
where $\bar{\kappa}=\frac{1}{2}(\kappa'+\kappa)$.

Adding the two together and putting $\kappa'=\kappa=0$ we get
\begin{eqnarray}
 M_{00}
&=&
 g_{PSS}g_{S}\,\left[\bar{u}(p's')u(ps)\right]
 \frac{1}{t-M_S^2+i\varepsilon}\ ,\label{sc4}
\end{eqnarray}
which is Feynman result \cite{henk1}.

In subsection \ref{ndepkad} we discussed the $n$-dependence of the
Kadyshevsky integral equation. In order to do that we need to know
the $n$-dependence of the amplitude \eqref{KIE12}
\begin{eqnarray}
 M_{0\kappa}^{(a+b)}
&=&
 M^{(a)}_{0\kappa}+M^{(b)}_{0\kappa}\ ,\nonumber\\
 \frac{\partial M_{0\kappa}^{(a+b)}}{\partial n^\beta}
&=&
 \kappa\ g_{PSS}g_{S}\,\left[\bar{u}(p's')u(ps)\right]\nonumber\\
&&
 \times\frac{n\cdot\Delta_t(\Delta_t)_\beta}{2A_t^3}\,
 \frac{(n\cdot\Delta_t)^2-3A_t^2-\frac{\kappa^2}{4}+2\kappa A_t}
 {\left((n\cdot\Delta_t)^2-\left(A_t-\frac{\kappa}{2}\right)^2+i\varepsilon\right)^2}
 \ .\label{sc5}
\end{eqnarray}
If we would only consider scalar meson exchange in the Kadyshevsky
integral equation the integrand would be of the form
\eqref{KIE13a}, where $h(\kappa)$ would by itself be of order
$O(\frac{1}{\kappa^2})$ as can be seen from \eqref{sc5}.
Therefore, the phenomenological "form factor" \eqref{KIE15} would
not be needed.\\

\noindent Since there's no propagator as far as Pomeron exchange
is concerned, the Kadyshevsky amplitude is the same as the Feynman
amplitude for Pomeron exchange \cite{henk1}
\begin{eqnarray}
 M_{\kappa'\kappa}&=&\frac{g_{PPP}g_P}{M}\,\left[\bar{u}(p's')u(p)\right]\ .\label{sc6}
\end{eqnarray}

\subsection{Vector Meson Exchange: Example}\label{ex1}

Before we go on with real vector meson exchange, we consider
simplified vector meson exchange. We use this as an example to
illustrate seaming problems that might occur in the results in the
Kadyshevsky formalism, especially when compared to those in the
Feynman formalism. We stress that although we consider the example
of simplified vector meson exchange, these peculiarities are
generally present when interaction Lagrangians containing
derivatives and/or higher spin fields ($s\geq1$) are considered.

In order to study simplified vector meson exchange we take
interaction Lagrangian \eqref{ing1b} and \eqref{ing2b}, without
the $\sigma_{\mu\nu}$-term
\begin{eqnarray}
 \mathcal{L}_I
&=&
 g\,\phi_a i\overleftrightarrow{\partial_\mu}\phi_b\cdot\phi^\mu
 +g\,\bar{\psi}\gamma_\mu\psi\cdot\phi^\mu\ ,\label{vb1}
\end{eqnarray}

\subsubsection{Naive Kadyshevsky Approach}\label{ex1kad}

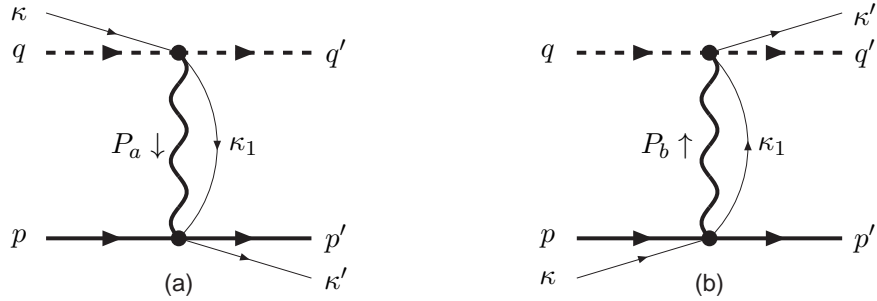
\begin{figure}[hbt]
\begin{center}
\begin{picture}(400,110)(0,0)
 \SetPFont{Helvetica}{9}
 \SetScale{1.0} \SetWidth{1.5}
 \DashArrowLine(50,90)(100,90){4}
 \DashArrowLine(100,90)(150,90){4}
 \ArrowLine(50,20)(100,20)
 \ArrowLine(100,20)(150,20)
 \Vertex(100,90){3}
 \Vertex(100,20){3}
 \Photon(100,20)(100,90){3}{3}

 \Text(85,55)[]{$P_a \downarrow$}
 \Text(125,55)[]{$\kappa_1$}
 \Text(40,20)[]{$p$}
 \Text(160,20)[]{$p'$}
 \Text(40,90)[]{$q$}
 \Text(160,90)[]{$q'$}

 \SetWidth{0.2}
 \ArrowLine(50,105)(100,90)
 \ArrowLine(100,20)(150,5)
 \ArrowArcn(65,55)(49.5,45,315)
 \PText(100,0)(0)[b]{(a)}
 \Text(40,105)[]{$\kappa$}
 \Text(160,5)[]{$\kappa'$}

 \SetScale{1.0} \SetWidth{1.5}
 \DashArrowLine(250,90)(300,90){4}
 \DashArrowLine(300,90)(350,90){4}
 \ArrowLine(250,20)(300,20)
 \ArrowLine(300,20)(350,20)
 \Vertex(300,90){3}
 \Vertex(300,20){3}
 \Photon(300,90)(300,20){3}{3}

 \Text(285,55)[]{$P_b \uparrow$}
 \Text(325,55)[]{$\kappa_1$}
 \Text(240,20)[]{$p$}
 \Text(360,20)[]{$p'$}
 \Text(240,90)[]{$q$}
 \Text(360,90)[]{$q'$}

 \SetWidth{0.2}
 \ArrowLine(250,5)(300,20)
 \ArrowLine(300,90)(350,105)
 \ArrowArc(265,55)(49.5,315,45)
 \PText(300,0)(0)[b]{(b)}
 \Text(240,5)[]{$\kappa$}
 \Text(360,105)[]{$\kappa'$}

\end{picture}
\end{center}
\caption{\sl Vector meson exchange in Kadyshevsky formalism.}
\label{fig:vectork}
\end{figure}

The Kadyshevsky diagrams for the (simplified) vector meson
exchange are shown in figure \ref{fig:vectork}.
For the various components of the diagrams we take the following
vertex functions
\begin{eqnarray}
 \Gamma^{\bar{\psi}\psi}_\mu
&=&
 g\,\gamma_\mu\ ,\nonumber\\
 \Gamma^{\phi\phi}_\mu
&=&
 g\,\left(q'+q\right)_\mu\ ,\label{vb2}
\end{eqnarray}
following from \eqref{vb1}, and the third line of \eqref{wick6}
for the propagator.

Applying the Kadyshevsky rules as given in appendix \ref{kadrules}
straightforwardly we get the following amplitudes
\begin{eqnarray}
 M^{(a,b)}_{\kappa'\kappa}
&=&
 -g^2\,\int\frac{d\kappa_1}{\kappa_1+i\varepsilon}\left[\bar{u}(p's')\gamma_\mu u(ps)\right]
 \left(g^{\mu\nu}-\frac{P_{a,b}^\mu P_{a,b}^\nu}{M_V^2}\right)\nonumber\\
&&
 \phantom{-g^2\,\int}\times
 \theta(P_{a,b}^0)\delta(P_{a,b}^2-M_V^2)\left(q'+q\right)_\nu\ ,\label{vb7}
\end{eqnarray}
The $\kappa_1$-integral is discussed in \eqref{sc2a} and
\eqref{sc2b}. We, therefore, give the results immediately
\begin{eqnarray}
 M^{(a)}_{\kappa'\kappa}
&=&
 -g^2\,\bar{u}(p's')\left[2\slQ-\frac{1}{M_V^2}
 \left((M_f-M_i)+\frac{1}{2}\,\sln(\kappa'-\kappa)-(\Delta_t\cdot n-A_t)\sln\right)
 \right.\nonumber\\*
&&
 \phantom{-g^2\,\bar{u}(p's')\left[\right.}\times
 \left(\frac{1}{4}\,(s_{p'q'}-s_{pq})+\frac{1}{4}\,(u_{pq'}-u_{p'q})-(m_f^2-m_i^2)
 \right.\nonumber\\
&&
 \phantom{-g^2\,\bar{u}(p's')\left[\right.\times\left(\right.}\left.\left.\vphantom{\frac{A}{A}}
 -2(\Delta_t\cdot n-A_t)n\cdot Q\right)\right]u(ps) \nonumber\\
&&
 \times\frac{1}{2A_t}\
 \frac{1}{\Delta_t\cdot n+\frac{1}{2}\left(\kappa'+\kappa\right)-A_t+i\varepsilon}
 \ ,\nonumber\\
 M^{(b)}_{\kappa'\kappa}
&=&
 -g^2\,\bar{u}(p's')\left[2\slQ-\frac{1}{M_V^2}
 \left((M_f-M_i)+\frac{1}{2}\,\sln(\kappa'-\kappa)-(\Delta_t\cdot n+A_t)\sln\right)\right.\nonumber\\
&&
 \phantom{-g^2\,\bar{u}(p's')\left[\right.}\times
 \left(\frac{1}{4}\,(s_{p'q'}-s_{pq})+\frac{1}{4}\,(u_{pq'}-u_{p'q})-(m_f^2-m_i^2)\right.
 \nonumber\\
&&
 \phantom{-g^2\,\bar{u}(p's')\left[\right.\left(\right.}\left.\left.\vphantom{\frac{A}{A}}
 -2(\Delta_t\cdot n+A_t)n\cdot Q\right)\right]u(ps)\nonumber\\
&&
 \times\frac{1}{2A_t}\
 \frac{1}{-\Delta_t\cdot n+\frac{1}{2}\left(\kappa'+\kappa\right)-A_t+i\varepsilon}
 \ .\label{vb11}
\end{eqnarray}
Adding the two together and putting $\kappa'=\kappa=0$ we should
get back the Feynman expression
\begin{eqnarray}
 M_{00}
&=&
 M^{(a)}_{00}+M^{(b)}_{00}\nonumber\\
&=&
 -g^2\bar{u}(p's')\left[2\slQ+\frac{(M_f-M_i)}{M_V^2}\,(m_f^2-m_i^2)\right]u(ps)\,\frac{1}{t-M_V^2+i\varepsilon}
 \nonumber\\
&&
 -g^2\bar{u}(p's')\left[\sln\right]u(ps)\,\frac{2Q\cdot n}{M_V^2}\ .\label{vb12}
\end{eqnarray}
From \eqref{vb12} we see that the first term on the rhs is indeed
the Feynman result. However, the second term on the rhs is an
unwanted, $n$-dependent, contact term.

As mentioned before, similar discrepancies are obtained when
couplings containing higher spin fields ($s\geq$1) are used.
Therefore, it seems that the Kadyshevsky formalism doesn't yield
the same results in these cases as the Feynman formalism when
$\kappa'$ and $\kappa$ are put to zero. Since the real difference
between Feynman formalism and Kadyshevsky formalism lies in the
treatment of the Time Ordered Product (TOP) or $\theta$-function
also the difference in results should find its origin in this
treatment.

In Feynman formalism derivatives are taken out of the TOP in order
to get Feynman functions, which may yield extra terms. This is
also the case in the above example
\begin{eqnarray}
 T[\phi^\mu(x)\phi^\nu(y)]
&=&
 -\left[g^{\mu\nu}+\frac{\partial^\mu\partial^\nu}{M_V^2}\right]i\Delta_F(x-y)
 -i\delta^\mu_0\delta^\nu_0\,\delta^4(x-y)\ ,\nonumber\\
 S_{fi}
&=&
 (-i)^2g^2\int d^4xd^4y\left[\bar{\psi}\gamma_\mu\psi\right]_x
 T[\phi^\mu(x)\phi^\nu(y)]
 \left[\phi_a\overleftrightarrow{i\partial_\nu}\phi_b\right]_y\
 ,\nonumber\\
 \Rightarrow M_{extra}
&=&
 -g^2\bar{u}(p's')\left[\sln\right]u(ps)\,\frac{2Q\cdot n}{M_V^2}
 \ .\label{vb13}
\end{eqnarray}
\footnote{If we include the $n^\mu$-vector in the
$\theta$-function of the TOP, which would not make a difference,
then we can make the replacement $\delta^\mu_0\rightarrow n^\mu$.
This, to make the result more general.} If we include the extra
term of \eqref{vb13} on the Feynman side we see that both
formalisms yield the same result.

Although we have exact equivalence between the two formalisms, the
result, though covariant, is still $n$-dependent, i.e.
frame-dependent. Of course this is not what we want. As it will
turn out there is another source of extra terms exactly cancelling
for instance the one that pops-up in our example (\eqref{vb12},
\eqref{vb13}). As mentioned in the introduction we present two
methods for getting these extra terms cancelling the one in
\eqref{vb12} and \eqref{vb13}: the TU method is more fundamental
and the GJ method is more systematic and pragmatic. Both methods
we will shortly introduce and apply to the problem in sections
\ref{ex2TU} and \ref{ex2GJ}, respectively.

\subsubsection{Takahahsi \& Umezawa Solution}\label{ex2TU}

In order to demonstrate the TU method \cite{Tak53a,Tak53b,Ume56}
we start with a rewritten version of the Yang-Feldman (YF)
equations \cite{Yang50} for a general interaction
\begin{eqnarray}
\mbox{\boldmath $\Phi$}_\alpha(x) &=& \Phi_\alpha(x) - \int d^4y\
 R_{\alpha\beta}(\partial)\ D_a(y)\ \Delta_{ret}(x-y)\cdot {\bf j}_{\beta;a}(y)\
 ,\label{eq:T.10}
\end{eqnarray}
where ${\bf \Phi}_\alpha(x)$ and $\Phi_\alpha(x)$ are fields in
the {\it Heisenberg Representation} (H.R.) and {\it Interaction
Representation} (I.R.), respectively. Furthermore, the vectors
$D_a(x)$ and ${\bf j}_{\alpha;a}(x)$ are defined to be
\begin{eqnarray}
 D_a(x) &\equiv&
 \left(1,\partial_{\mu_1},\partial_{\mu_1}\partial_{\mu_2},\ldots\right)\ ,
 \nonumber\\
 {\bf j}_{\alpha;a}(x)
&\equiv &
 \left(-\frac{\partial{\cal L}_I}{\partial\mbox{\boldmath $\Phi$}_{\alpha}(x)}\ ,\
 -\frac{\partial{\cal L}_I}{\partial\left(\partial_{\mu_1}\mbox{\boldmath $\Phi$}_{\alpha}(x)\right)}\ ,\
 -\frac{\partial{\cal L}_I}{\partial\left(\partial_{\mu_1}\partial_{\mu_2}\mbox{\boldmath $\Phi$}_{\alpha}(x)\right)}\ ,\
 \ldots\right)\ ,
 \label{eq:T.11a}
\end{eqnarray}
Next, a free auxiliary field $\Phi_{\alpha}(x,\sigma)$ is
introduced, where $\sigma$ is a space-like surface and $x$ does
not necessarily lie on $\sigma$. We pose that it has the following
form
\begin{equation}
 \Phi_\alpha(x,\sigma) \equiv \Phi_\alpha(x) + \int^\sigma_{-\infty} d^4y\
 R_{\alpha\beta}(\partial)D_a(y)\ \Delta(x-y)\cdot {\bf j}_{\beta;a}(y)\ ,
\label{eq:T.12}
\end{equation}
Combining \eqref{eq:T.12} with \eqref{eq:T.10} leads to
\begin{eqnarray}
 \mbox{\boldmath $\Phi$}_\alpha(x)
&=&
 \Phi_\alpha(x/\sigma)
 +\frac{1}{2}\int d^4y\left[\vphantom{\frac{A}{A}}
 R_{\alpha\beta}(\partial)D_a(y),\epsilon(x-y)\right]
 \Delta(x-y)\cdot {\bf j}_{\beta;a}(y)\ .\nonumber\\\label{eq:T.16}
\end{eqnarray}
This equation will be used to express the fields in the H.R. in
terms of fields in the I.R.

In appendix \ref{BMP} it is explained that the auxiliary fields
and the fields in the I.R. are related by a unitary operator using
the BMP theory. Also it is shown how the interaction Hamiltonian
should be deduced.\\

\noindent Applying these concepts to our example we determine the
"currents" via \eqref{eq:T.11a}
\begin{eqnarray}
 \mbox{\boldmath $j$}_{\phi_a,a}
&=&
 \left(-g\ i\partial_\mu\mbox{\boldmath $\phi$}_b\cdot\mbox{\boldmath $\phi$}^\mu,
 ig\ \mbox{\boldmath $\phi$}_b\cdot\mbox{\boldmath $\phi$}^\mu \right)\ ,\nonumber\\
 \mbox{\boldmath $j$}_{\phi_b,a}
&=&
 \left(g\ i\partial_\mu\mbox{\boldmath $\phi$}_a\cdot\mbox{\boldmath $\phi$}^\mu,
 -ig\ \mbox{\boldmath $\phi$}_a\cdot\mbox{\boldmath $\phi$}^\mu \right)\ ,\nonumber\\
 \mbox{\boldmath $j$}_{\psi,a}
&=&
 \left(-g\ \gamma_\mu\mbox{\boldmath $\psi$}
 \cdot\mbox{\boldmath $\phi$}^\mu,0\right)\ ,\nonumber\\
 \mbox{\boldmath $j$}_{\phi^\mu,a}
&=&
 \left(-g\ \mbox{\boldmath $\phi$}_a\overleftrightarrow{i\partial_\mu}\mbox{\boldmath
 $\phi$}_b
 -g\ \mbox{\boldmath $\bar{\psi}$}\gamma_\mu\mbox{\boldmath $\psi$},0\right)\ .\label{ex2.1}
\end{eqnarray}
Using \eqref{eq:T.16} we can express the fields in the H.R. in
terms of fields in the I.R., i.e. free fields
\begin{eqnarray}
 \mbox{\boldmath $\phi$}_a(x)
&=&
 \phi_a(x/\sigma)\ ,\nonumber\\
 \mbox{\boldmath $\phi$}_b(x)
&=&
 \phi_b(x/\sigma)\ ,\nonumber\\
 \partial_\mu \mbox{\boldmath $\phi$}_a(x)
&=&
 \left[\partial_\mu \phi_a(x,\sigma)\right]_{x/\sigma}
 +\frac{1}{2}\,\int d^4y\left[\partial_\mu^x\partial_\nu^y,\epsilon(x-y)\right]\Delta(x-y)
 \left(ig\phi_b\cdot\phi^\nu\right)_y\nonumber\\
&=&
 \left[\partial_\mu \phi_a(x,\sigma)\right]_{x/\sigma}
 +ign_\mu\phi_b\ n\cdot\phi\ ,\nonumber\\
 \partial_\mu \mbox{\boldmath $\phi$}_b(x)
&=&
 \left[\partial_\mu \phi_b(x,\sigma)\right]_{x/\sigma}
 +\frac{1}{2}\,\int d^4y\left[\partial_\mu^x\partial_\nu^y,\epsilon(x-y)\right]\Delta(x-y)
 \left(-ig\phi_a\cdot\phi^\nu\right)_y\nonumber\\
&=&
 \left[\partial_\mu \phi_b(x,\sigma)\right]_{x/\sigma}
 -ign_\mu\phi_a\ n\cdot\phi\ ,\nonumber\\
 \mbox{\boldmath $\psi$}(x)
&=&
 \psi(x/\sigma)\ ,\nonumber\\
 \mbox{\boldmath $\phi$}^\mu(x)
&=&
 \phi^\mu(x/\sigma)
 +\frac{1}{2}\,\int d^4y\left[\left(-g^{\mu\nu}-\frac{\partial^\mu\partial^\nu}{M_V^2}\right),\epsilon(x-y)\right]\Delta(x-y)
 \nonumber\\
&&
 \phantom{\phi^\mu(x/\sigma)+\frac{1}{2}\,\int}\times
 \left(-g\phi_a\overleftrightarrow{i\partial_\nu}\phi_b-g\bar{\psi}\gamma_\nu\psi\right)_y\nonumber\\
&=&
 \phi^\mu(x/\sigma)-\frac{g\,n^\mu}{M_V^2}
 \left(\phi_a n\cdot\overleftrightarrow{i\partial}\phi_b+\bar{\psi}\sln\psi\right)\ .\label{ex2.2}
\end{eqnarray}
As can be seen from \eqref{eq:T.16} the first term on the rhs is a
free field and the second term contains the current expressed in
terms of fields in the H.R., which on their turn are expanded
similarly. Therefore, one gets coupled equations. In solving these
equations we assumed that the coupling constant is small and
therefore considered only terms up to first order in the coupling
constant in the expansion of the fields in the H.R. Practically
speaking, the currents on the rhs of \eqref{ex2.2} are expressed
in terms of free fields.

These expansions \eqref{ex2.2} are used in the commutation
relations of the fields with the interaction Hamiltonian
(\eqref{eq:T.23} of appendix \ref{BMP})
\begin{eqnarray}
 \left[\phi_a(x),\mathcal{H}_I(y)\right]
&=&
 iU^{-1}(\sigma)\Delta(x-y)\left[-g\ i\partial_\mu\mbox{\boldmath $\phi$}_b\cdot\mbox{\boldmath $\phi$}^\mu
 +g\ \overleftarrow{i\partial_\mu}\mbox{\boldmath $\phi$}_b\cdot\mbox{\boldmath $\phi$}^\mu\right]_yU(\sigma)
 \nonumber\\
&=&
 i\Delta(x-y)\left[ -g\ \overleftrightarrow{i\partial_\mu}\phi_b\cdot\phi^\mu
 \right.\nonumber\\
&&
 \left.
 +\frac{g^2}{M_V^2}\ n\cdot\overleftrightarrow{i\partial}\phi_b
 \left(\phi_an\cdot\overleftrightarrow{i\partial}\phi_b+\bar{\psi}\sln\psi\right)
 -g^2\ \phi_a(n\cdot\phi)^2
 \vphantom{\frac{A}{A}}\right]_y\nonumber\\
 \left[\psi(x),\mathcal{H}_I(y)\right]
&=&
 iU^{-1}(\sigma)(i\slpart+M)\Delta(x-y)\left[-g\ \gamma_\mu\mbox{\boldmath $\psi$}
 \cdot\mbox{\boldmath $\phi$}^\mu\right]_yU(\sigma)\nonumber\\*
&=&
 i(i\slpart+M)\Delta(x-y)\nonumber\\*
&&
 \times\left[-g\ \gamma_\mu\psi\cdot\phi^\mu
 +\frac{g^2}{M_V^2}\ \sln\psi
 \left(\phi_a n\cdot\overleftrightarrow{i\partial}\phi_b+\bar{\psi}\sln\psi\right)
 \right]_y\ ,\nonumber\\
 \left[\phi^\mu(x),\mathcal{H}_I(y)\right]
&=&
 iU^{-1}(\sigma)\left(-g^{\mu\nu}-\frac{\partial^\mu\partial^\nu}{M_V^2}\right)\Delta(x-y)
 \nonumber\\
&&
 \times\left[-g\ \mbox{\boldmath $\phi$}_a\overleftrightarrow{i\partial_\nu}\mbox{\boldmath
 $\phi$}_b
 -g\ \mbox{\boldmath $\bar{\psi}$}\gamma_\nu\mbox{\boldmath $\psi$}\right]_yU(\sigma)\nonumber\\
&=&
 i\left(-g^{\mu\nu}-\frac{\partial^\mu\partial^\nu}{M_V^2}\right)\Delta(x-y)
 \left[-g\ \phi_a\overleftrightarrow{i\partial_\nu}\phi_b-g\ \bar{\psi}\gamma_\nu\psi
 \right.\nonumber\\
&&
 \left.\vphantom{\frac{a}{a}}
 -g^2\,n_\nu\,\phi^2_an\cdot\phi-g^2\,n_\nu\,\phi^2_bn\cdot\phi\right]_y\ .\label{ex2.3}
\end{eqnarray}
As stated below \eqref{eq:T.23} these are the fundamental
equations from which the interaction Hamiltonian can be determined
\begin{eqnarray}
 \mathcal{H}_I
&=&
 -g\ \phi_a\overleftrightarrow{i\partial_\mu}\phi_b\cdot\phi^\mu-g\ \bar{\psi}\gamma_\mu\psi\cdot\phi^\mu
 -\frac{g^2}{2}\ \phi_a^2(n\cdot\phi)^2-\frac{g^2}{2}\ \phi_b^2(n\cdot\phi)^2
 \nonumber\\
&&
 +\frac{g^2}{2M_V^2}\ \left[\bar{\psi}\sln\psi\right]^2
 +\frac{g^2}{M_V^2}\ \left[\bar{\psi}\sln\psi\right]\left[\phi_a n\cdot\overleftrightarrow{i\partial}\phi_b\right]
 +\frac{g^2}{2M_V^2}\ \left[\phi_a n\cdot\overleftrightarrow{i\partial}\phi_b\right]^2
 \nonumber\\
&&
 +O(g^3)\ldots\ .\label{ex2.4}
\end{eqnarray}
If equation \eqref{ex2.2} was solved completely, then the rhs of
\eqref{ex2.2} would contain higher orders in the coupling constant
and therefore also the interaction Hamiltonian \eqref{ex2.4}.
These terms are indicated by the ellipsis.

If we want to include the external quasi fields as in appendix
\ref{secquant}, then the easy way to do this is to apply
\eqref{a12} straightforwardly. However, since we want to derive
the interaction Hamiltonian from the interaction Lagrangian we
would have to include a $\bar{\chi}(x)\chi(x)$ pair in \eqref{vb1}
similar to \eqref{a12}. This would mean that the terms of order
$g^2$ in \eqref{ex2.4} are quartic in the quasi field, where two
of them can be contracted
\begin{eqnarray}
 \bar{\chi}(x)\bcontraction{}{\chi}{(x)}{\bar{\chi}}\chi(x)\bar{\chi}(x)\chi(x)
 =\bar{\chi}(x)\theta[n(x-x)]\chi(x)\ .\label{ex2.4a}
\end{eqnarray}
Defining the $\theta$-function to be 1 in its origin we assure
that all terms in the interaction Hamiltonian \eqref{ex2.4}
relevant to $\pi N$-scattering are quadratic in the external quasi
fields, even higher order terms in the coupling constant.

The only term of order $g^2$ in \eqref{ex2.4} that gives a
contribution to the first order in the S-matrix describing $\pi
N$-scattering is the second term on the second line in the rhs of
\eqref{ex2.4}. Its contribution to the first order in the S-matrix
is
\begin{eqnarray}
 S^{(1)}_{fi}
&=&
 -i\int d^4x\mathcal{H}_I(x)
 =\frac{-ig^2}{M_V^2}\int d^4x
 \left[\bar{\psi}\sln\psi\right]\left[\phi_a
 n\cdot\overleftrightarrow{i\partial}\phi_b\right]_x\nonumber\\
&=&
 \frac{-ig^2}{M_V^2}\,\bar{u}(p's')\sln u(ps) n\cdot(q'+q)\
 ,\nonumber\\
 \Rightarrow M_{canc}
&=&
 g^2\,\bar{u}(p's')\sln u(ps)\frac{2n\cdot Q}{M_V^2}\ .\label{ex2.5}
\end{eqnarray}
Indeed we see that this term \eqref{ex2.5} cancels the extra term
in \eqref{vb12}.

From \eqref{ex2.4} one can see that the interaction Hamiltonian
contains not only terms of order $g$, but also higher order terms.
In our example we see that the $g^2$ terms in the interaction
Hamiltonian is responsible for the cancellation. In this light we
would also like to mention the specific example of scalar
electrodynamics as described in \cite{itzyk}, section 6-1-4. There
the interaction Hamiltonian also contains a term of order $g^2$,
which has the same purpose as in our case. The method described in
\cite{itzyk} is not generally applicable, whereas the above
described method, although applied to a specific example, is.

\subsubsection{Gross \& Jackiw Solution}\label{ex2GJ}

The essence of the Gross and Jackiw method \cite{Gross69} is to
define a different TOP: the $T^*$ product, which is by definition
$n$-independent
\begin{eqnarray}
 T^*(x,y)&=& T(x,y;n)+\tau(x,y;n)\ ,
\end{eqnarray}
Studying the $n$-dependence is done in the same way as described
in subsection \ref{ndepkad}
\begin{eqnarray}
 P^{\alpha\beta}\frac{\delta}{\delta n^\beta} T^*(x,y)
&=&
 P^{\alpha\beta}\frac{\delta}{\delta n^\beta}T(x,y;n)
 +P^{\alpha\beta}\frac{\delta}{\delta n^\beta}\tau(x,y; n)\equiv0\ . \quad\label{GJ3}
\end{eqnarray}

In our applications we are interested in second order
contributions to $\pi N$-scattering. Therefore, we analyze the
$n$-dependence of the TOP of two interaction Hamiltonians, where
we take it to be just $\mathcal{H}_I=-\mathcal{L}_I$
\begin{eqnarray}
 P^{\alpha\beta}\frac{\delta}{\delta n^\beta}T(x,y;n)
&=&
 P^{\alpha\beta}(x-y)_\beta \delta\left[n\cdot(x-y)\right]\
 \left[{\cal H}_I(x),{\cal H}_I(y)\right]\ .\qquad\label{GJ5}
\end{eqnarray}
In general one has for equal time commutation relations
\begin{eqnarray}
 \delta[n(x-y)]\left[{\cal H}_I(x),{\cal H}_I(y)\right]
&=&
 \left[C(n)+P^{\alpha\beta}S_\alpha(n)\partial_\beta\right.\nonumber\\
&&
 \left.\ +P^{\alpha\beta}P^{\mu\nu}Q_{\alpha\mu}(n)\partial_\beta\partial_\nu
 +\ldots\right]\delta^4(x-y)\ .\label{GJ7}
\end{eqnarray}
where the ellipsis stand for higher order derivatives. We will
only consider (and encounter) up to quadratic derivatives. The
$S^\alpha$ and $Q^{\alpha\beta}$ terms in \eqref{GJ7} are known in
the literature as {\it Schwinger terms}.

It should be mentioned that in \cite{Gross69} only the first two
terms on the rhs of \eqref{GJ7} are considered.

Using the fact that the TOP and therefore also the $T^*$ product
appears in the S-matrix as an integrand, we are allowed to use
partial integration for the $S_\alpha(n)$ and $Q_{\alpha\beta}(n)$
terms. The $C(n)$ always vanishes. Furthermore, we use the fact
that $P^{\alpha\beta}$ is a projection operator. With these
considerations we find from \eqref{GJ3}-\eqref{GJ7} the extra
terms
\begin{eqnarray}
 \tau(x-y;n)=\int^n dn'^\beta
 \left[S_\beta(n')+P^{\mu\nu}\left(Q_{\beta\mu}(n')+Q_{\mu\beta}(n')
 \vphantom{\frac{a}{a}}\right)\partial_\nu\right]
 \delta^4(x-y)\ .\label{GJ9}
\end{eqnarray}
In principle the rhs of \eqref{GJ9} can also contain a constant
term, i.e. independent of $n^\mu$. But since we are looking for
$n^\mu$-dependent terms only, this term is irrelevant.

Now, we're going to apply the method of Gross and Jackiw. The
"covariantized" equal time commutator of interaction Hamiltonians
is
\begin{eqnarray}
&&
 \delta[n(x-y)]\left[\mathcal{H}_I(x),\mathcal{H}_I(y)\right]=\nonumber\\
&=&
 g^2\left\{\frac{1}{M_V^2}\left(
 \left[\psi\sln\psi\right]_x\left[\phi_a\overleftrightarrow{i\partial_\mu}\phi_b\right]_y
 +\left[\psi n_\mu\psi\right]_x\left[\phi_an\cdot\overleftrightarrow{i\partial}\phi_b\right]_y
 \right.\right.\nonumber\\
&&
 \phantom{g^2\{\frac{1}{M_V^2}\left(\right.}
 +\left[\phi_an\cdot\overleftrightarrow{i\partial}\phi_b\right]_x\left[\psi n_\mu\psi\right]_y
 +\left[\phi_a\overleftrightarrow{i\partial_\mu}\phi_b\right]_x\left[\psi\sln\psi\right]_y
 \nonumber\\
&&
 \phantom{g^2\{\frac{1}{M_V^2}\left(\right.}
 +\left[\psi\sln\psi\right]_y\left[\psi\gamma_\mu\psi\right]_x
 +\left[\psi\gamma_\mu\psi\right]_y\left[\psi\sln\psi\right]_x
 \nonumber\\
&&
 \phantom{g^2\{\frac{1}{M_V^2}\left(\right.}\left.
 +\left[\phi_an\cdot\overleftrightarrow{i\partial}\phi_b\right]_y
 \left[\phi_a\overleftrightarrow{i\partial_\mu}\phi_b\right]_x
 +\left[\phi_a\overleftrightarrow{i\partial_\mu}\phi_b\right]_y
 \left[\phi_an\cdot\overleftrightarrow{i\partial}\phi_b\right]_x\right)
 \nonumber\\
&&
 \phantom{g^2\left\{\right.}
 +\phi_a(y)n\cdot\phi(x)\phi_a(x)\phi_\mu(y)+\phi_a(y)\phi_\mu(x)\phi_a(x)n\cdot\phi(y)
 \nonumber\\
&&
 \phantom{g^2\left\{\right.}\left.
 +\left[\phi_bn\cdot\phi\right]_x\left[\phi_b\phi_\mu\right]_y
 +\left[\phi_b\phi_\mu\right]_x\left[\phi_bn\cdot\phi\right]_y
 \vphantom{\frac{A}{A}}\right\}\,P^{\mu\rho}i\partial_\rho\delta^4(x-y)\
 .\label{ex2.8}
\end{eqnarray}
Comparing this with \eqref{GJ7} we see that the terms between
curly brackets coincide with $-iS_\alpha(n)$; the
$Q_{\alpha\beta}(n)$ terms are absent. Therefore, the
$\tau$-function, representing the compensating terms, becomes by
means of \eqref{GJ9} and \eqref{ex2.8}
\begin{eqnarray}
 \tau(x-y;n)
&=&
 ig^2\left[\frac{1}{M_V^2}\left(
 2\left[\psi\sln\psi\right]\left[\phi_a n\cdot\overleftrightarrow{i\partial}\phi_b\right]
 +\left[\psi\sln\psi\right]^2+\left[\phi_a n\cdot\overleftrightarrow{i\partial}\phi_b\right]^2\right)
 \right.\nonumber\\
&&
 \phantom{ig^2[}\left.
 +\phi_a^2(n\cdot\phi)^2+\phi_b^2(n\cdot\phi)^2
 \vphantom{\frac{A}{A}}\right]\delta^4(x-y)\ .\label{ex2.9}
\end{eqnarray}
Its contribution to $\pi N$-scattering S-matrix and amplitude is
\begin{eqnarray}
 S^{(2)}_{canc}
&=&
 \frac{(-i)^2}{2!}\,\int d^4xd^4y\ \frac{2ig^2}{M_V^2}\
 \left[\psi\sln\psi\right]\left[\phi_a n\cdot\overleftrightarrow{i\partial}\phi_b)\right]
 \delta^4(x-y)\ ,\nonumber\\
 M_{canc}
&=&
 g^2\ \bar{u}(p's')\sln u(ps)\frac{2n\cdot Q}{M_V^2}\ ,\label{ex2.10}
\end{eqnarray}
which is the same expression as the cancelling amplitude derived
from the TU scheme in \eqref{ex2.5}.

\subsubsection{$\bar{P}$ Approach}\label{Pbar}

From the forgoing subsections (sections \ref{ex2GJ} and
\ref{ex2TU}) we have seen that if we add all contributions,
results in the Feynman formalism and in the Kadyshevsky formalism
are the same (of course we need to put $\kappa'=\kappa=0$). Also,
we have seen from \eqref{vb13} and the forgoing subsections that
if we bring out the derivatives out of the TOP in Feynman
formalism not only do we get Feynman functions, but also the
$n$-dependent contact terms cancel out.

Unfortunately, this is not the case in Kadyshevsky formalism.
There, all $n$-dependent contact terms cancel out after adding up
the amplitudes. So, when calculating an amplitude according to the
Kadyshevsky rules in appendix \ref{kadrules} one always has to
keep in mind the contributions as described in sections
\ref{ex2TU} and \ref{ex2GJ}. For practical purposes this is not
very convenient.

Inspired by the Feynman procedure we could also do the same in
Kadyshevsky formalism, namely let the derivatives not only act on
the vector meson propagator \footnote{With 'propagator' we mean
the $\Delta^+(x-y)$ and not the Feynman propagator
$\Delta_F(x-y)$.} but also on the quasi particle propagator
($\theta$-function). In doing so, we know that all contact terms
cancel out; just as in Feynman formalism.

We show the above in formula form.
\begin{eqnarray}
&&
 \theta[n(x-y)]\partial^\mu_x\partial^\nu_x\Delta^{(+)}(x-y)
 +\theta[n(y-x)]\partial^\mu_x\partial^\nu_x\Delta^{(+)}(y-x)
 \nonumber\\
&=&
 \partial^\mu_x\partial^\nu_x\theta[n(x-y)]\Delta^{(+)}(x-y)
 +\partial^\mu_x\partial^\nu_x\theta[n(y-x)]\Delta^{(+)}(y-x)
 \nonumber\\
&&
 +in^\mu n^\nu\delta^4(x-y)\nonumber\\
&=&
 \frac{i}{2\pi}\int\frac{d\kappa_1}{\kappa_1+i\varepsilon}\,\int\frac{d^4P}{(2\pi)^3}\,\theta(P^0)\delta(P^2-M_V^2)
 \left(-\bar{P}_\mu\bar{P}_\nu\right)\nonumber\\
&&
 \times\left(e^{-i\kappa_1n(x-y)}e^{-iP(x-y)}+e^{i\kappa_1n(x-y)}e^{iP(x-y)}\right)
 \nonumber\\
&&
 +in^\mu n^\nu\delta^4(x-y)\ ,\label{ex2.11}
\end{eqnarray}
where $\bar{P}=P+n\kappa_1$. In this way the second order in the
S-matrix becomes
\begin{eqnarray}
 S^{(2)}_{fi}
&=&
 -g^2{\int}d^4xd^4y\left[\bar{u}(p's')\gamma_{\mu}u(ps)\right]\left(q'+q\right)_{\nu}e^{-ix(q-q')}e^{iy(p'-p)}
 \nonumber\\
&&
 \times\frac{i}{2\pi}\int\frac{d\kappa_1}{\kappa_1+i\varepsilon}\,\int\frac{d^4P}{(2\pi)^3}
 \theta(P^0)\delta(P^2-M_V^2)
 \left(-g^{\mu\nu}+\frac{\bar{P}^{\mu}\bar{P}^{\nu}}{M_V^2}\right)
 \nonumber\\
&&
 \times\left(e^{-i\kappa_1n(x-y)}e^{-iP(x-y)}e^{in\kappa'x-in\kappa y}
 +e^{i\kappa_1n(x-y)}e^{iP(x-y)}e^{-in\kappa'x+in\kappa y}\right)
 \nonumber\\
&&
 +ig^2{\int}d^4x\left[\bar{u}(p's')\sln u(ps)\right]n\cdot\left(q'+q\right)e^{-ix(q-q'-p'+p-n\kappa'+n\kappa)}
 \ .\qquad\label{ex2.12}
\end{eqnarray}
We see that the second term on the rhs of \eqref{ex2.12} brings
about an amplitude, which is exactly the same as in \eqref{vb12}
and \eqref{vb13} and is to be cancelled by \eqref{ex2.5} and
\eqref{ex2.10}.

Performing the various integrals correctly we get
\begin{eqnarray}
(a)&\Rightarrow& \left\{
\begin{array}{ccc}
 \kappa_1&=&\Delta_t\cdot n-A_t+\frac{1}{2}\left(\kappa'+\kappa\right)n \\
 \bar{P} &=&\Delta_t+\frac{1}{2}\left(\kappa'+\kappa\right)n \\
\end{array}\right.\nonumber\\
(b)&\Rightarrow& \left\{
\begin{array}{ccc}
 \kappa_1&=&-\Delta_t\cdot n-A_t+\frac{1}{2}\left(\kappa'+\kappa\right)n \\
 \bar{P} &=&-\Delta_t+\frac{1}{2}\left(\kappa'+\kappa\right)n \\
\end{array}\right.\ .\label{ex2.13}
\end{eqnarray}
This yields for the invariant amplitudes
\begin{eqnarray}
 M^{(a)}_{\kappa'\kappa}
&=&
 -g^2\ \bar{u}(p's')\left[\vphantom{\frac{A}{A}}2\slQ
 +\frac{1}{M_V^2}\left((M_f-M_i)+\frac{1}{2}(\kappa'-\kappa)\sln+\sln\bar{\kappa}\right)
 \right.\nonumber\\
&&
 \times\left.\left(\left(m_f^2-m_i^2\right)+\frac{1}{4}\left(s_{pq}-s_{p'q'}+u_{p'q}-u_{pq'}\right)
 +2\bar{\kappa}Q\cdot n\right)\right]u(ps)\nonumber\\*
&&
 \times\frac{1}{2A_t}\ \frac{1}{\Delta_t\cdot n+\bar{\kappa}-A_t+i\varepsilon}\ ,\nonumber\\
 M^{(b)}_{\kappa'\kappa}
&=&
 -g^2\ \bar{u}(p's')\left[\vphantom{\frac{A}{A}}2\slQ
 +\frac{1}{M_V^2}\left((M_f-M_i)+\frac{1}{2}(\kappa'-\kappa)\sln-\sln\bar{\kappa}\right)
 \right.\nonumber\\
&&
 \times\left.\left(\left(m_f^2-m_i^2\right)+\frac{1}{4}\left(s_{pq}-s_{p'q'}+u_{p'q}-u_{pq'}\right)
 -2\bar{\kappa}Q\cdot n\right)\right]u(ps)\nonumber\\
&&
 \times\frac{1}{2A_t}\ \frac{1}{-\Delta_t\cdot n+\bar{\kappa}-A_t+i\varepsilon}
 \nonumber\\
 M
&=&
 M^{(a)}_{00}+M^{(b)}_{00}\nonumber\\
&=&
 -g^2\bar{u}(p's')\left[2\slQ+\frac{\left(M_f-M_i\right)}{M_V^2}
 \left(m_f^2-m_i^2\right)\right]u(ps)\frac{1}{t-M_V^2+i\varepsilon}
 \ ,\label{ex2.14}
\end{eqnarray}
where $\bar{\kappa}=\frac{1}{2}\left(\kappa'+\kappa\right)$. As
before we get back the Feynman expression for the amplitude if we
add both amplitudes obtained in Kadyshevsky formalism and put
$\kappa'=\kappa=0$. The big advantage of this procedure is that we
do not need to worry about the contribution $n$-dependent contact
terms because they cancelled out when introducing $\bar{P}$.

It should be noticed however that the $\bar{P}$-method is only
possible when both Kadyshevsky contributions at second order are
added. This becomes clear when looking at the first two lines of
\eqref{ex2.11}: Letting the derivatives also act on the
$\theta$-function gives compensating terms for the
$\Delta^{(+)}(x-y)$-part and for the $\Delta^{(-)}(x-y)$-part.
Only when added together they combine to the $\delta^4(x-y)$-part.

Also it becomes clear from \eqref{ex2.11} that at least two
derivatives are needed to generate the $\delta^4(x-y)$-part.
Therefore, when there's only one derivative, for instance in the
case of baryon exchange (so, no derivatives in coupling only in
the propagator) at second order, the $\delta^4(x-y)$-part is not
present and it is not necessary to use the $\bar{P}$-method. In
these cases it doesn't matter for the summed diagrams whether or
not the $\bar{P}$-method is used, however for the individual
diagrams it does make a difference. This ambiguity is absent in
Feynman theory, there derivatives are always taken out of the TOP
(which is similar to the $\bar{P}$-method, as discussed above) in
order to come to Feynman propagators.

In the forgoing we have demonstrated the $\bar{P}$-method for
simplified vector meson exchange and strictly speaking for
$\kappa'=\kappa=0$. We stress, however, that this method is
generally applicable, i.e. for $\kappa',\kappa\neq 0$ and for
general couplings containing multiple derivatives and/or higher
spin fields.

\subsection{Real Vector Meson Exchange}\label{sectionvme}

Now that we have discussed how to deal with multiple derivatives
and/or higher spin fields in the Kadyshevsky formalism by means of
the simplified vector meson exchange example, we're prepared to
deal with real vector meson exchange. In order to do so we use the
interaction Lagrangians as in \eqref{ing1b} and \eqref{ing2b}.
From these interaction Lagrangians we distillate the already
exposed vertex function in \eqref{vb2} (second line) and
\begin{eqnarray}
  \Gamma_{VNN}^\mu
&=&
 g_{V}\,\gamma^{\mu}
 +\frac{f_V}{2M_V}\,\left(p'-p\right)_{\alpha}\sigma^{\alpha\mu}
 \ .\label{v1}
\end{eqnarray}
The Kadyshevsky diagrams representing vector meson exchange are
already exposed in figure \ref{fig:vectork}. Applying the
Kadyshevsky rules of appendix \ref{kadrules} and the $\bar{P}$
method described in section \ref{Pbar} we obtain the following
amplitudes
\begin{eqnarray}
 M^{(a)}_{\kappa'\kappa}
&=&
 -g_{VPP}\,\bar{u}(p's')\left[\vphantom{\frac{A}{A}}2g_V\slQ\right.\nonumber\\
&&
 -\frac{g_V}{M_V^2}\ \kappa'\sln \left(\frac{1}{4}\left(s_{p'q'}-s_{pq}+u_{pq'}-u_{p'q}\right)
 +2\bar{\kappa}Q\cdot n\right)
 \nonumber\\
&&
 +\frac{f_V}{2M_V}\left(4M\slQ+\frac{1}{2}\left(u_{pq'}+u_{p'q}\right)
 -\frac{1}{2}\left(s_{p'q'}+s_{pq}\right)\right.\nonumber\\
&&
 \phantom{\frac{f_V}{2M_V}(}
 -\frac{1}{M_V^2}\left(M^2+m^2-\frac{1}{2}\left(\frac{1}{2}(t_{p'p}+t_{q'q})+u_{pq'}+s_{pq}\right)\right.\nonumber\\
&&
 \phantom{\frac{f_V}{2M_V}(-\frac{1}{M_V^2}(}\left.
 +2M\sln\kappa'+\frac{1}{4}\left(\kappa'-\kappa\right)^2
 -\left(p'+p\right)\cdot n\bar{\kappa}\vphantom{\frac{A}{A}}\right)
 \nonumber\\
&&
  \phantom{\frac{f_V}{2M_V}(-(}\left.\left.\times
 \left(\frac{1}{4}(s_{p'q'}-s_{pq})+\frac{1}{4}(u_{pq'}-u_{p'q})+2\bar{\kappa}n\cdot Q\right)
 \right)\right]u(ps)\nonumber\\
&&
 \times\frac{1}{2A_t}\ \frac{1}{\Delta_t\cdot n+\bar{\kappa}-A_t+i\varepsilon}\ ,\nonumber\\
 M^{(b)}_{\kappa'\kappa}
&=&
 -g_{VPP}\,\bar{u}(p's')\left[\vphantom{\frac{A}{A}}2g_V\slQ\right.\nonumber\\
&&
 +\frac{g_V}{M_V^2}\ \kappa\sln\left(\frac{1}{4}\left(s_{p'q'}-s_{pq}+u_{pq'}-u_{p'q}\right)-2\bar{\kappa}Q\cdot n\right)
 \nonumber\\
&&
 +\frac{f_V}{2M_V}\left(4M\slQ+\frac{1}{2}\left(u_{pq'}+u_{p'q}\right)
 -\frac{1}{2}\left(s_{p'q'}+s_{pq}\right)\right.\nonumber\\
&&
 \phantom{\frac{f_V}{2M_V}(}
 -\frac{1}{M_V^2}\left(M^2+m^2-\frac{1}{2}\left(\frac{1}{2}(t_{p'p}+t_{q'q})+u_{pq'}+s_{pq}\right)\right.\nonumber\\
&&
 \phantom{\frac{f_V}{2M_V}(-\frac{1}{M_V^2}(}\left.
 -2M\sln\kappa+\frac{1}{4}\left(\kappa'-\kappa\right)^2
 +\left(p'+p\right)\cdot n\bar{\kappa}\vphantom{\frac{A}{A}}\right)
 \nonumber\\
&&
  \phantom{\frac{f_V}{2M_V}(-(}\left.\left.\times
 \left(\frac{1}{4}(s_{p'q'}-s_{pq})+\frac{1}{4}(u_{pq'}-u_{p'q})-2\bar{\kappa}n\cdot Q\right)
 \right)\right]u(ps)\nonumber\\
&&
 \times\frac{1}{2A_t}\ \frac{1}{-\Delta_t\cdot n+\bar{\kappa}-A_t+i\varepsilon}\ .\label{v2}
\end{eqnarray}
The sum of the two in the limit of $\kappa'=\kappa=0$ yields
\begin{eqnarray}
 M_{00}
&=&
 -g_{VPP}\ \bar{u}(p's')\left[\vphantom{\frac{A}{A}}2g_V\slQ
 +\frac{f_V}{2M_V}\left(\vphantom{\frac{a}{a}}
 (u-s)+4M\slQ\right)\right]u(ps)\nonumber\\
&&
 \phantom{-}
 \times\frac{1}{t-M_V^2+i\varepsilon}\ ,\label{v3}
\end{eqnarray}
which is, again, the Feynman result \cite{henk1}.

Just as in section \ref{sectionsme} we study the $n$-dependence of
the amplitude. This, in light of the $n$-dependence of the
Kadyshevsky integral equation (see section \ref{ndepkad}).
\begin{eqnarray}
 M_{0\kappa}^{(a+b)}
&=&
 M^{(a)}_{0\kappa}+M^{(b)}_{0\kappa},\nonumber\\*
&=&
 -g_{VPP}\ \bar{u}(ps)\left[2g_V\slQ +\frac{f_V}{2M_V}\left(4M\slQ+\frac{1}{2}\left(u_{pq'}+u_{p'q}\right)
 \right.\right.\nonumber\\
&&
 \hspace{1cm}\left.\left.
 -\frac{1}{2}\left(s_{p'q'}+s_{pq}\right)\right)\right]u(ps)
 \frac{A_t-\frac{\kappa}{2}}{A_t}
 \frac{1}{(\Delta_t\cdot n)^2-\left(A_t-\frac{\kappa}{2}\right)^2+i\varepsilon}\nonumber\\
&&
 -\frac{g_V f_V\kappa}{2M_V^3}\ \bar{u}(ps)\left[
 \frac{1}{2}(p'+p)\cdot n (Q\cdot n)\kappa\left(A_t-\frac{\kappa}{2}\right)\right.\nonumber\\
&&
 \hspace{1cm}
 +\frac{1}{8}(p'+p)\cdot n\left(s_{p'q'}-s_{pq}+u_{pq'}-u_{p'q}\right)\Delta_t\cdot n
 \nonumber\\
&&
 \hspace{1cm}
 -n\cdot Q \left(M^2+m^2-\frac{1}{2}\left(\frac{1}{2}(t_{p'p}+t_{q'q})+u_{pq'}+s_{pq}\right)+\frac{\kappa^2}{4}\right)
 \nonumber\\
&&
 \left.\hspace{1cm}\times\vphantom{\frac{A}{A}}
 \Delta_t\cdot n\right]u(ps)\,
 \frac{1}{A_t}\,\frac{1}{(\Delta_t\cdot n)^2-(\frac{\kappa}{2}-A_t)^2+i\varepsilon}
 \nonumber\\
&&
 +\frac{g_{VPP}\kappa}{M_V^2}\ \bar{u}(p's')\left[\sln
 \left(g_V+\frac{f_VM}{M_V}\right)\left(\frac{1}{4}(s_{p'q'}-s_{pq}+u_{pq'}-u_{p'q})\vphantom{\frac{A}{A}}
 \right.\right.\nonumber\\
&&
 \hspace{1cm}\left.\left.\vphantom{\frac{A}{A}}
 +\kappa n\cdot Q\right)\right]u(ps)\,
 \frac{1}{2A_t}\ \frac{1}{\Delta_t\cdot n+\frac{\kappa}{2}-A_t+i\varepsilon}\ .\label{v4}
\end{eqnarray}
Differentiating this with respect to $n^\alpha$ in the same way as
in \eqref{sc5} we know that the result will contain an overall
factor of $\kappa$. This can be seen as follows: The first term in
\eqref{v4} is very similar to $M^{(a+b)}_{0\kappa}$ in
\eqref{sc5}. Therefore, the overall factor of $\kappa$ when
differentiating with respect to $n^\alpha$ is obvious. All other
terms in \eqref{v4} contain already an overall factor of $\kappa$,
which doesn't change when differentiating.

As can be seen from \eqref{v4} the numerator is of higher degree
in $\kappa$ then the denominator. Therefore, the function
$h(\kappa)$ in \eqref{KIE13a} will not be of order
$O(\frac{1}{\kappa^2})$ and the "form factor" \eqref{KIE15} is
necessary.

In \eqref{v2} as well as in \eqref{sc3} we have taken $u$ and
$\bar{u}$ spinors. The reason behind this is pair suppression
which we will discuss in paper II.

\begin{appendices}

\section{Kadyshevsky Rules}\label{kadrules}

Just as in Feynman theory Kadyshevsky amplitudes can be
represented by Kadyshevsky diagrams. Since the basic starting
points are the same as in Feynman theory we take a general Feynman
diagram and give the Kadyshevsky rules from there on to construct
the amplitude $M_{fi}$. Here, we define the amplitude as
\begin{eqnarray}
 S_{fi}=\delta_{fi}-i(2\pi)^2\delta^4\left(P_f-P_i\right)\,M_{fi}\ ,\label{defM}
\end{eqnarray}
where $P_{f/i}$ is the sum of the final/initial momenta.
\vspace{0.5cm}

\underline{\bf Kadyshevsky Rules:} \vspace{0.5cm}

\noindent{\bf 1)} Arbitrarily number the vertices of the diagram.
\vspace{0.5cm}

\noindent{\bf 2)} Connect the vertices with a quasi particle line,
assigned to it a momentum $n\kappa_s$ ($s=1\ldots n-1$). Attach to
vertex $1$ an incoming initial quasi particle with momentum
$n\kappa$ and attach to vertex $n$ an outgoing final quasi
particle with momentum $n\kappa'$ \footnote{Obviously these quasi
particle may not appear as initial or final states, since they are
not physical particles. However, since we use Kadyshevsky diagrams
as input for an integral equation we allow for external quasi
particles.}. \vspace{0.5cm}

\noindent{\bf 3)} Orient each internal momentum such that it
leaves a vertex with a lower number than the vertex it enters. If
2 fermion lines with opposite momentum direction come together in
one vertex assign a $+$ symbol to one line and a $-$ to the other.
Each possibility to do this yields a different Kadyshevsky
diagram. \vspace{0.5cm}

\noindent{\bf 4)} Assign to each internal quasi particle line a
propagator $\frac{1}{\kappa_s+i\varepsilon}$. \vspace{0.5cm}

\noindent{\bf 5)} Assign to all other internal lines the
appropriate Wightman function of \eqref{wick6}. Assign to a
fermion line with a $\pm$ symbol: $S^{(\pm)}(P)$ (see {\bf 3}))
\begin{eqnarray}
 \Delta^{(+)}(P)
&=&
 \theta(P^0)\delta(P^2-M^2)\ ,\nonumber\\
 S^{(\pm)}(P)
&=&
 \Lambda^{(1/2)}(\pm P)\ \theta(P^0)\delta(P^2-M^2)\
 ,\nonumber\\
 \Delta_{\mu\nu}^{(+)}(P)
&=&
 \Lambda^{(1)}_{\mu\nu}(P)\ \theta(P^0)\delta(P^2-M^2)\
 ,\nonumber\\
 S^{(\pm)}_{\mu\nu}(P)
&=&
 \Lambda^{(3/2)}_{\mu\nu}(\pm P)\ \theta(P^0)\delta(P^2-M^2)\ ,\label{wick6}
\end{eqnarray}
where
\begin{eqnarray}
 \Lambda^{(1/2)}_{\mu\nu}(P)
&=&
 \left(\slP + M\right)\ ,\nonumber\\
 \Lambda^{(1)}_{\mu\nu}(P)
&=&
 \left(-g_{\mu\nu}+\frac{P_\mu P_\nu}{M^2}\right)\ ,\nonumber\\
 \Lambda^{(3/2)}_{\mu\nu}(P)
&=&
 -\left(\slP + M\right)
 \left(g_{\mu\nu}-\frac{1}{3}\,\gamma_\mu\gamma_\nu
 -\frac{2P_\mu P_\nu}{3M^2}\right.\nonumber\\
&&
 \phantom{-\left(\slP + M\right)(}\left.
 +\frac{1}{3M}\left(P_\mu\gamma_\nu-\gamma_\mu P_\nu\right)\right)\ .\label{wick5}
\end{eqnarray}
\vspace{0.5cm}

\noindent{\bf 6)} There's momentum conservation at the vertices,
including the quasi particle momenta. \vspace{0.5cm}

\noindent{\bf 7)} Integrate over the internal quasi momenta:
$\int_{-\infty}^\infty d\kappa_s$. \vspace{0.5cm}

\noindent{\bf 8)} Integrate over those internal momenta not fixed
by momentum conservation at the vertices:
$\int_{-\infty}^{\infty}\frac{d^4P}{(2\pi)^3}$. \vspace{0.5cm}

\noindent{\bf 9)} Include a $-$ sign for every fermion loop.
\vspace{0.5cm}

\noindent{\bf 10)} Include a $-$ sign for identical initial or
final fermions. \vspace{0.5cm}

\noindent{\bf 11)} Repeat the various steps for all different
numberings in {\bf 1}. \vspace{0.5cm}

It is clear from {\bf 3)} and {\bf 11)} that one Feynman diagram
leads to several Kadyshevsky diagrams. Generally, one Feynman
diagram leads to $n!$ Kadyshevsky diagrams, where $n$ is the
number of vertices (or; the order). Especially for higher order
diagrams this leads to a dramatic increase of labour. Fortunately,
we will only consider second order diagrams.

A few remarks need to be made about these rules as far as the
choice of definition is concerned. In {\bf 3)} we have followed
\cite{Kad64} to orient the internal momenta. Furthermore we have
chosen to use the integral representation of the $\theta$-function
\begin{eqnarray}
 \theta[n\cdot(x-y)]
&=&
 \frac{i}{2\pi}\int d\kappa_1
 \frac{e^{-i\kappa_1n\cdot(x-y)}}{\kappa_1+i\varepsilon}
 \ ,\label{a5}
\end{eqnarray}
instead of its complex conjugate. Since the $\theta$-function is
real, this is also a proper representation, originally used in the
papers of Kadyshevsky. To understand why we have chosen to deviate
from the original approach, consider the S-matrix
\begin{eqnarray}
 S
&=&
 1+\sum_{n=1}^{\infty}\left(-i\right)^n\int_{-\infty}^{\infty}
 d^4x_1\dots d^4x_n\
 \theta[n(x_1-x_{2})]\ldots\theta[n(x_{n-1}-x_n)]
 \nonumber\\
&&
 \phantom{1+\sum_{n=1}^{\infty}\left(-i\right)^n\int_{-\infty}^{\infty}}
 \times\mathcal{H}_{I}(x_1)\ldots\mathcal{H}_{I}(x_n)\ .\qquad\label{smatrix}
\end{eqnarray}
In each order $S_n$ there is a factor $(-i)^n$ already in the
definition. In that specific order there are $(n-1)$
$\theta$-functions, each containing a factor $i$ from the integral
representation \eqref{a5}. Therefore, every $S_n$ will, regardless
the order, contain a factor $(-i)$. Hence, the amplitude $M_{fi}$,
defined in \eqref{defM}, will not contain overall factors of $i$,
anymore.

The momentum space $S^{(-)}(P)$-functions differ an overall minus
sign by their coordinate space analogs $\langle
0|\bar{\psi}(x)\psi(y)|0\rangle=S^{(-)}(x-y)$. The reason for that
is twofold. In many cases the Wightman functions $S^{(-)}(x-y)$,
including the overall minus sign, appear in combination with the
Normal Ordered Product (NOP):
$N(\psi\bar{\psi})=-N(\bar{\psi}\psi)$. Therefore, the minus signs
cancel. In all other cases the Wightman functions $S^{(-)}(x-y)$
appear in fermion loops and are therefore responsible for the
fermion loop minus sign in {\bf 9)}, since every fermion loop will
contain an odd number of $S^{(-)}(x-y)$ functions. We stress that
this method of defining the Kadyshevsky rules for fermions differs
from the original one in \cite{Kad68}.

\section{Second Quantization}\label{secquant}

When discussing the Kadyshevsky rules in subsection \ref{kadrules}
and the Kadyshevsky integral equation in \eqref{KIE7} we allowed
for quasi particles to occur in the initial and final state. In
order to do this properly a new theory needs to be set up
containing quasi particle creation and annihilation operators. It
is set up in such a way that external quasi particles occur in the
S-matrix as trivial exponentials so that when the external quasi
momenta are taken to be zero the Feynman expression is obtained.
We, therefore, require that the vacuum expectation value of the
quasi particles is the $\theta$-function
\begin{eqnarray}
 <0|\chi(nx)\bar{\chi}(nx')|0>=\theta[n(x-x')]\ ,\label{a6}
\end{eqnarray}
and that a quasi field operator acting on a state with quasi
momentum $(n)\kappa$ only yields a trivial exponential
\begin{eqnarray}
 \chi(nx)|\kappa>&=&e^{-i\kappa nx}\ ,\nonumber\\
 <\kappa|\bar{\chi}(nx)&=&e^{i\kappa nx}\ .\label{a7}
\end{eqnarray}
Assuming that a state with quasi momentum $(n)\kappa$ is created
in the usual way
\begin{eqnarray}
 a^\dagger(\kappa)|0>&=&|\kappa>\ ,\nonumber\\
 <0|a(\kappa)&=&<\kappa|\ ,\label{a8}
\end{eqnarray}
we have from the requirements \eqref{a6} and \eqref{a7} the
following momentum expansion of the fields
\begin{eqnarray}
 \chi(nx)&=&\frac{i}{2\pi}\int\frac{d\kappa}{\kappa+i\varepsilon}
            \ e^{-i\kappa nx}a(\kappa)\ ,\nonumber\\
 \bar{\chi}(nx')&=&\frac{i}{2\pi}\int\frac{d\kappa}{\kappa+i\varepsilon}
            \ e^{i\kappa nx'}a^\dagger(\kappa)\ ,\label{a9}
\end{eqnarray}
and the fundamental commutation relation of the creation and
annihilation operators
\begin{eqnarray}
 \left[a(\kappa),a^\dagger(\kappa')\right]=-i2\pi\kappa\delta(\kappa-\kappa')\
 .\label{a10}
\end{eqnarray}
From this commutator \eqref{a10} it is clear that the quasi
particle is not a physical particle nor a ghost.

Now that we have set up the second quantization for the quasi
particles we need to include them in the S-matrix. This is done by
redefining it
\begin{eqnarray}
 S&=&1+\sum_{n=1}(-i)^n\int d^4x_1\ldots d^4x_n
 \mathcal{\tilde{H}}_I(x_1)\ldots\mathcal{\tilde{H}}_I(x_n)\
 ,\label{a11}
\end{eqnarray}
where
\begin{eqnarray}
 \mathcal{\tilde{H}}_{I}(x)&\equiv&\mathcal{H}_{I}(x)\bar{\chi}(nx)\chi(nx)
 \ .\label{a12}
\end{eqnarray}
In this sense contraction of the quasi fields causes propagation
of this field between vertices, just as in the Feynman formalism.
Those quasi particles that are not contracted are used to
annihilate external quasi particles form the vacuum.
\begin{eqnarray}
&&
 S^{(2)}(p's'q'n\kappa';psqn\kappa)=\nonumber\\*
&=&
 (-i)^2\,\int d^4x_1d^4x_2
 <\pi N\chi|\mathcal{\tilde{H}}_{I}(x_1)\mathcal{\tilde{H}}_{I}(x_2)|\pi
 N\chi>\nonumber\\
&=&
 (-i)^2\,\int d^4x_1d^4x_2
 <0|b(p's')a(q')a(\kappa')\nonumber\\
&&
 \times
 \left[\bar\chi(nx_1)\mathcal{H}_{I}(x_1)\chi(nx_1)\bar\chi(nx_2)\mathcal{H}_{I}(x_2)\chi(nx_2)
 \vphantom{\frac{A}{A}}\right]a^\dagger(\kappa)a^\dagger(q)b^\dagger(ps)|0>
 \nonumber\\
&=&
 (-i)^2\,\int d^4x_1d^4x_2
 \ e^{in\kappa'x_1}e^{-in\kappa x_2}\nonumber\\
&&
 \times
 <0|b(p's')a(q')\mathcal{H}_{I}(x_1)\theta[n(x_1-x_2)]\mathcal{H}_{I}(x_2)a^\dagger(q)b^\dagger(ps)|0>
 \ .\label{a13}
\end{eqnarray}
For the $\pi$ and $N$ fields we use the well-known momentum
expansion
\begin{eqnarray}
 \phi(x)&=&\int\frac{d^3l}{(2\pi)^32E_{l}}\left[a(l)e^{-ilx}+a^\dagger(l)e^{ilx}\right]\ ,\nonumber\\
 \psi(x)&=&\sum_{r}\int\frac{d^3k}{(2\pi)^32E_{k}}
 \left[b(k,r)u(k,r)e^{-ikx}+d^\dagger(k,r)v(k,r)e^{ikx}\right]\
 ,\qquad\label{a14}
\end{eqnarray}
where the creation and annihilation operators satisfy the
following (anti-) commutation relations
\begin{eqnarray}
 [a(k),a^\dagger(l)]
&=&
 (2\pi)^3\,2E_k\,\delta^3(k-l)\ ,\nonumber\\
 \{b(k,s),b^\dagger(l,r)\}
&=&
 (2\pi)^3\,2E_k\,\delta_{sr}\delta^3(k-l)=\{d(k,s),d^\dagger(l,r)\}
 \ .\label{a15}
\end{eqnarray}
Putting $\kappa'=\kappa=0$ in \eqref{a13} we see that we get the
second order in the S-matrix expansion for $\pi N$-scattering as
in Feynman formalism. Of course this is what we required from the
beginning: external quasi particle momenta only occur in the
S-matrix as exponentials.

So, we know now how to include the external quasi particles in the
S-matrix and therefore we also know what their effect is on
amplitudes. For practical purposes we will not use the S-matrix as
in \eqref{a11}, but keep the above in mind. In those cases where
the (possible) inclusion of external quasi fields is less trivial
we will make some comments.

\section{BMP Theory}\label{BMP}

According to Haag's theorem \cite{haag55} in general there does
not exist a unitary transformation which relates the fields in the
I.R. and the fields in the H.R. On the other hand there is no
objection against the existence of an unitary $U[\sigma]$ relating
the TU-auxiliary fields and the fields in the I.R.
\begin{equation}
 \Phi_\alpha(x,\sigma) = U^{-1}[\sigma]\ \Phi_\alpha(x)\ U[\sigma]\
 .\label{BMP1}
\end{equation}
Here, we follow the framework of Bogoliubov and collaborators
\cite{BMP58,BS59,BLT75}, to which we refer to as the BMP theory,
to prove \eqref{BMP1} in a straightforward way (see appendix
\ref{app:U.e}).

The BMP theory was originally constructed to bypass the use of an
unitary operator $U$ as a mediator between the fields in the H.R.
and in the I.R.

\subsection{Set-up}\label{app:U.b}

In the description of the BMP theory we will only consider scalar
fields. By the assumption of asymptotic completeness the S-matrix
is taken to be a functional of the asymptotic fields
$\phi_{as,\rho}(x)$, where $as=in,out$. In the following we use
$in$-fields, i.e. $\phi_\rho(x)= \phi_{in,\rho}(x)$
\begin{eqnarray}
 S
&=&
 1+\sum_{n=1}^\infty\int d^4x_1 \ldots d^4x_n\ S_n(x_1\alpha_1,\ldots , x_n\alpha_n)\cdot
 \nonumber\\
&&
 \phantom{1+\sum_{n=1}^\infty\int}\times
 :\phi_{\alpha_1}(x_1) \ldots \phi_{\alpha_n}(x_n) :\ .
\label{eq:U.11}
\end{eqnarray}
Here, concepts like unitarity and the stability of the vacuum,
i.e. $\langle 0| S | 0\rangle=1$, and the 1-particle states, i.e.
$\langle 0| S | 1\rangle=0$ are assumed. The {\it Heisenberg
current}, i.e. the current in the H.R., is defined as
\footnote{Note that in \cite{BLT75} the out-field is used. Then
\begin{eqnarray*}
 {\bf J}_\rho(x) = i \frac{\delta S}{\delta\phi_\rho(x)} S^\dagger\ .
\end{eqnarray*}
Also, we take a minus sign in the definition of the current.}
\begin{equation}
 {\bf J}_\rho(x) = -i S^\dagger\ \frac{\delta
 S}{\delta\phi_\rho(x)}\ .\label{eq:U.14}
\end{equation}
We note that for a hermitean field $\phi_\rho(x)$ the current is
also hermitean, due to unitarity. {\it Microcausality} takes the
form, see \cite{BS59}, section 17 \footnote{ Here $x \leq y$ means
either $(x-y)^2 \geq 0$ and $x^0 < y^0$ or $(x-y)^2 < 0$. So, the
point $x$ is in the past of or is spacelike separated from the
point $y$. },
\begin{equation}
 \frac{\delta {\bf J}_\rho(x)}{\delta\phi_\lambda(y)} = 0\ \ ,\ \ {\rm for}\ \
 x \leq y\ .\label{eq:U.15}
\end{equation}

It can be shown that the notion of microcausality is reflected in
the expression of the S-matrix as the Time-Ordered exponential.
See \cite{BS59} for the details on this point of view. It can also
be shown that with the current (\ref{eq:U.14}) the asymptotic
fields $\phi_{in/out,\rho}(x)$ satisfy a YF type of equation (as
in \eqref{eq:U.3})
\begin{eqnarray}
 \mbox{\boldmath $\phi$}_\rho(x) &=& \phi_{in/put,\rho}(x)+\int d^4y\,
 \Delta_{ret/adv}(x-y)\, {\bf J}_{\rho}(y)\ ,
\end{eqnarray}
giving the Heisenberg fields $\mbox{\boldmath$\phi$}_\rho(x)$ in
terms of the $\phi_{in/out}(x)$fields.\\

\noindent Lehmann, Symanzik, and Zimmermann (LSZ) \cite{LSZ55}
formulated an asymptotic condition utilizing the notion of weak
convergence in the Hilbert space of state vectors. See e.g.
\cite{Bjorken} for an detailed exposition of the LSZ-formalism.
The correspondence of BMP theory with LSZ is obtained by the
identification
\begin{equation}
 {\bf J}_\rho(x) = -i S^\dagger\ \frac{\delta S}{\delta\phi_\rho(x)}
 \equiv \left(\Box + m^2\right)\ \mbox{\boldmath $\phi$}_\rho(x)\
 .\label{eq:U.21}
\end{equation}
As is explained in for instance \cite{BLT75}, the local
commutivity of the currents follows from microcausality
\eqref{eq:U.15}. Using the YF equations one can show that for
space-like separations the fields in the H.R. commute with the
currents and among themselves, as was assumed in the
LSZ-formalism. For more details and results of BMP see
\cite{BMP58,BS59,BLT75}.

\subsection{Application to Takahashi-Umezawa scheme}\label{app:U.e}

In this subsection we introduce the auxiliary field similar to
\eqref{eq:T.12}
\begin{equation}
 \phi(x,\sigma) \equiv \phi(x) - \int_{-\infty}^\sigma d^4x'\ \Delta(x-x')\
 {\bf J}(x')\ ,
\label{eq:U.51}
\end{equation}
and prove that $\phi(x)$ and $\phi(x,\sigma)$ satisfy the same
(usual) commutation relations. Such a relation justifies the
existence of an unitary operator connecting the two as in
\eqref{BMP1}.

The difference of the commutation relations is, using
(\ref{eq:U.51}),
\begin{eqnarray}
&&
 \left[\vphantom{\frac{A}{A}} \phi(x,\sigma),\phi(y,\sigma)\right] -
 \left[\vphantom{\frac{A}{A}} \phi(x), \phi(y)\right]\nonumber\\
&=&
 -\int_{-\infty}^\sigma d^4y'\ \Delta(y-y')
 \left[\vphantom{\frac{A}{A}} \phi(x), {\bf J}(y')\right]
 +\int_{-\infty}^\sigma d^4x'\ \Delta(x-x')
 \left[\vphantom{\frac{A}{A}} \phi(y), {\bf J}(x')\right] \nonumber\\
&&
 + \int_{-\infty}^\sigma\int_{-\infty}^\sigma d^4x' d^4y'\
 \Delta(x-x')\Delta(y-y')
 \left[\vphantom{\frac{A}{A}} {\bf J}(x'), {\bf J}(y')\right]\ .\label{eq:U.53}
\end{eqnarray}
Since the $S$-operator is an expansion in asymptotic fields, so is
${\bf J}(x)$ by means of its definition in terms of this
$S$-operator \eqref{eq:U.14}. Now, from the commutation relations
of the asymptotic fields one has
\begin{equation}
 \left[\vphantom{\frac{A}{A}} \phi_\rho(x), {\bf J}_\sigma(y)\right]
 = i \int d^4x'\ \Delta(x-x') \frac{\delta {\bf J}_\sigma(y)}{\delta\phi_\rho(x')}\ . \label{eq:U.28}
\end{equation}
Using this in \eqref{eq:U.53} we have
\begin{eqnarray}
&&
 \left[\vphantom{\frac{A}{A}} \phi(x,\sigma),\phi(y,\sigma)\right] -
 \left[\vphantom{\frac{A}{A}} \phi(x), \phi(y)\right]\nonumber\\
&=&
 -i \int_{-\infty}^\sigma d^4y'\ \int_{-\infty}^\infty d^4x' \Delta(x-x')\Delta(y-y')
 \frac{\delta {\bf J}(y')}{\delta \phi(x')}
 \nonumber\\
&&
 +i \int_{-\infty}^\sigma d^4x'\ \int_{-\infty}^\infty d^4y'\ \Delta(x-x')\Delta(y-y')
 \frac{\delta {\bf J}(x')}{\delta \phi(y')} \nonumber\\
&&
 -i \int_{-\infty}^\sigma d^4x'\ \int_{-\infty}^\sigma d^4y'\ \Delta(x-x')\Delta(y-y')
 \left( \frac{\delta {\bf J}(x')}{\delta \phi(y')} -\frac{\delta {\bf J}(y')}{\delta \phi(x')}\right)
 \nonumber\\
&=&
 0\ .\label{eq:U.54}
\end{eqnarray}
Cancellation takes place in \eqref{eq:U.54} when the second
integral of the first two term on the rhs in \eqref{eq:U.54} is
split up:
$\int_{-\infty}^{\infty}=\int_{-\infty}^{\sigma}+\int_{\sigma}^{\infty}$.
The remaining terms are zero because of the microcausility
condition \eqref{eq:U.15}. Although we shown the proof for scalar
fields only, the generalization to other types of fields is easily
made.

Complementary to what is in \cite{Tak53a,Tak53b,Ume56} we
explicitly show that the unitary operator in \eqref{BMP1} is not
any operator but the one connected to the S-matrix. We, therefore,
consider (general) $in$- and $out$-fields. Their relation to the
fields in the H.R. is
\begin{eqnarray}
 \mbox{\boldmath $\Phi$}_\alpha(x)
&=&
 \Phi_{in ,\alpha}(x) + \int d^4y\ R_{\alpha\beta}(\partial)\ \Delta_{ret}(x-y)\ {\bf J}_\beta(y) \nonumber\\
&=&
 \Phi_{out,\alpha}(x) + \int d^4y\ R_{\alpha\beta}(\partial)\ \Delta_{adv}(x-y)\ {\bf J}_\beta(y)\ ,
\label{eq:U.3}
\end{eqnarray}
where $\Delta_{ret}(x-y)=-\theta(x^0-y^0)\Delta(x-y)$ and
$\Delta_{adv}(x-y)=\theta(y^0-x^0)\Delta(x-y)$.

Equation \eqref{eq:U.3} makes clear that the choice of the Green
function determines the choice of the free field ($in$- or
$out$-field) to be used. In this light we make the following
identification: $\Phi_\alpha(x,-\infty)\equiv\Phi_{in,\alpha}(x)$,
since we have used the retarded Green function in section
\ref{ex2TU} (text below \eqref{eq:T.10}). With \eqref{eq:U.3} we
can also relate the $out$-field to the auxiliary field
$\Phi_\alpha(x,\infty)=\Phi_{out,\alpha}(x)$.

Using these identifications in \eqref{BMP1} we obtain the relation
between $\Phi_{\alpha,in}(x)$ and $\Phi_{\alpha,out}(x)$
\begin{eqnarray}
 \Phi_{\alpha,in}(x)
&=&
 U^{-1}[-\infty]U[\infty]\ \Phi_{\alpha,out}(x)\ U^{-1}[\infty]U[-\infty]
 \nonumber\\
 \Phi_{in,\alpha}(x)
&=&
 S\Phi_{out,\alpha}S^{-1}\ .\label{in2}
\end{eqnarray}
Obviously, the operator connecting the $in$- and $out$-fields is
the S-matrix, where the relation between $U[\sigma]$ and the
S-matrix is
\begin{eqnarray}
 U[\sigma]&=&T\left[exp\left(-i\int_{-\infty}^{\sigma}d^4x\mathcal{H}_I(x)\right)\right]\ ,\nonumber\\
 U[\infty]&=&S\ ,\qquad U[-\infty]=1\ .\label{in3}
\end{eqnarray}
To make contact with the interaction Hamiltonian we follow
\cite{Tak53a,Tak53b,Ume56} for completion by realizing that the
unitary operator satisfies the Tomonaga-Schwinger equation
\begin{equation}
 i\frac{\delta U[\sigma]}{\delta\sigma(x)} = \left.\vphantom{\frac{A}{A}}
 {\cal H}_I(x;n) U[\sigma]\right|_{x/\sigma} = U[\sigma]\ {\cal H}_I(x/\sigma;n)\
 .\label{eq:T.20}
\end{equation}
Here, the interaction Hamiltonian will in general depend on the
vector $n_\mu(x)$ locally normal to the surface $\sigma(x)$, i.e.
$n^\mu(x)d\sigma_\mu=0$. It is hermitean because of the unitarity
of $U[\sigma]$. Then, from \eqref{BMP1} and \eqref{eq:T.20} one
gets that
\begin{equation}
 i \frac{\delta \Phi_\alpha(x,\sigma)}{\delta\sigma(y)} = U^{-1}[\sigma]
 \left[\vphantom{\frac{A}{A}} \Phi_\alpha(x) , \mathcal{H}_I(y;n) \right]\
 U[\sigma]\ .\label{eq:T.21}
\end{equation}
On the other hand, varying \eqref{eq:T.12} with respect to
$\sigma(y)$ gives
\begin{equation}
 i\frac{\delta \Phi_\alpha(x,\sigma)}{\delta\sigma(y)} =
 i\  D_a(y)\ R_{\alpha\beta}(\partial)\ \Delta(x-y)\cdot {\bf j}_{\beta;a}(y)\ .
\label{eq:T.22}
\end{equation}
Comparing (\ref{eq:T.21}) and (\ref{eq:T.22}) gives the relation
\begin{eqnarray}
 \left[\vphantom{\frac{A}{A}} \Phi_\alpha(x) , \mathcal{H}_I(y;n) \right]
&=&
 i\  U[\sigma]\left[D_a(y)\ R_{\alpha\beta}(\partial)\ \Delta(x-y)\cdot{\bf j}_{\beta;a}(y)
 \vphantom{\frac{A}{A}}\right] U^{-1}[\sigma]\ .\nonumber\\\label{eq:T.23}
\end{eqnarray}
{\it This is the fundamental equation by which the interaction
Hamiltonian must be determined}.

\section{Remarks on the Haag Theorem}\label{haag}

Here, we take a closer look at the connection between the fields
in the H.R. and in the I.R. in the covariant formulation of
Tomonaga and Schwinger \cite{Tom46,Schw48}
\begin{eqnarray}
 \mbox{\boldmath $\Phi$}_\alpha(x)
&=&
 U^{-1}[\sigma]\ \Phi_\alpha(x)\ U[\sigma]\ , \label{eq:T.1}
\end{eqnarray}
This in light of the Haag theorem \cite{haag55}, which states that
if there is an unitary operator connecting fields in two
representations at some time (as in \eqref{eq:T.1}), where the
field in one representation is free, both fields are free. This
would lead to a triviality.

The question is whether this situation \eqref{eq:T.1} is
applicable to our case. In order to answer that question we look
at the results of the previous subsection (appendix \ref{BMP}). By
introducing the auxiliary field in the scalar case as in
\eqref{eq:T.12} (or for general fields as in \eqref{eq:U.51}) we
proved \eqref{BMP1} using BMP theory.

Now, we start with \eqref{eq:T.10} and use similar arguments to
come to
\begin{eqnarray}
 \mbox{\boldmath $\Phi$}_\alpha(x)
&=&
 \Phi_\alpha(x) + \int_{-\infty}^{\infty} d^4y\
 D_a(y)\ R_{\alpha\beta}(\partial)\ \theta[n(x-y)]\Delta(x-y)\cdot {\bf
 j}_{\beta;a}(y)\nonumber\\
&=&
 \Phi_\alpha(x) + \int_{-\infty}^{\infty} d^4y\ \theta[n(x-y)]\
 D_a(y)\ R_{\alpha\beta}(\partial)\ \Delta(x-y)\cdot {\bf
 j}_{\beta;a}(y)\nonumber\\
&&
 + \int_{-\infty}^{\infty} d^4y \left[D_a(y)\ R_{\alpha\beta}(\partial), \theta[n(x-y)]\right]
 \Delta(x-y)\cdot {\bf j}_{\beta;a}(y)\ ,\nonumber\\
 \Rightarrow \mbox{\boldmath $\Phi$}_\alpha(x)
&=&
 U^{-1}[\sigma]\ \Phi_\alpha(x)\ U[\sigma]|_{x/\sigma}\nonumber\\
&&
 + \frac{1}{2}\int_{-\infty}^{\infty} d^4y \left[D_a(y)\ R_{\alpha\beta}(\partial), \epsilon(x-y)\right]
 \Delta(x-y)\cdot {\bf j}_{\beta;a}(y)\ .\label{haag2}
\end{eqnarray}
\footnote{We have included the $n^\mu$-vector in the first line of
\eqref{haag2}, which causes no effect. The reason for this
inclusion is that we can keep the surface $\sigma$ general, though
space-like.} The above is different from what is exposed in
\cite{Bjorken} (ch 17.2). The difference is the commutator part in
\eqref{haag2} and this term is non-zero for theories with
couplings containing derivatives and higher spin fields, carefully
excluded in the treatment of \cite{Bjorken}. Therefore
\eqref{haag2} could be seen as an extension of what is written in
\cite{Bjorken}.

Returning to Haag's theorem we see that if the last term in
\eqref{haag2} is absent there is an unitary operator connecting
$\mbox{\boldmath $\Phi$}_\alpha(x)$ and $\Phi_\alpha(x)$ and
therefore they are both free fields. Such theories can then be
considered as trivial, although they can still be useful as
effective theories.

In our application we use various interaction Lagrangians (for the
overview see section \ref{OBE}) to be used in order to describe
the various exchange (and resonance (paper II)) processes. Whether
or not the non-vanishing commutator part in \eqref{haag2} is
present depends on the process under consideration. In the vector
meson exchange diagrams (section \ref{sectionvme}) and in the
spin-3/2 exchange and resonance diagrams (paper II) those
commutator parts are non-vanishing. If we include pair suppression
in the way we do in paper II also in the spin-1/2 exchange and
resonance diagrams the commutator parts will be non-vanishing. So,
if we take the model as a whole (all diagrams) then it is most
certainly not trivial in the sense of the Haag theorem.

\end{appendices}

\end{document}